\documentclass[letterpaper, reprint, superscriptaddress, amsmath, amssymb, aps, prl, nofootinbib]{revtex4-2}

\makeatletter

\renewcommand\onecolumngrid{%
  \do@columngrid{one}{\@ne}%
  \def\set@footnotewidth{\onecolumngrid}%
  \def\footnoterule{\kern-6pt\hrule width 1.5in\kern6pt}%
}

\renewcommand\twocolumngrid{%
  \def\footnoterule{%
    \dimen@\skip\footins
    \divide\dimen@\thr@@
    \kern-\dimen@\hrule width.5in\kern\dimen@
  }%
  \do@columngrid{mlt}{\tw@}%
}

\makeatother

\usepackage{mathtools}
\usepackage[english]{babel}
\usepackage[utf8x]{inputenc}
\usepackage{amsmath}
\usepackage{amsfonts}
\usepackage{amssymb}
\usepackage{graphicx}
\usepackage[margin=1in]{geometry}
\usepackage{float}
\usepackage{indentfirst}
\usepackage{color} 
\usepackage{xcolor}
\usepackage{subfigure}
\usepackage{ulem}
\usepackage{comment}
\graphicspath{ {Images/} }

\usepackage{titlesec}
\titleformat{\section}[runin]
  {\normalfont\itshape} 
  {}        
  {0em}                
  {}                   
  [---]                
\titlespacing*{\section}
  {0pt}                
  {3.5ex plus 1ex minus .2ex} 
  {0em}                

\newcommand{\rfone}{r_1^f}
\newcommand{\rftwo}{r_2^f}
\newcommand{\rione}{r_1^i}
\newcommand{\ritwo}{r_2^i}

\newcommand{\epsmax}{\epsilon_{\textrm{max}}}
\newcommand{\Tcoll}{T_{\textrm{coll}}}
\newcommand{\Torb}{T_{\textrm{orb}}}
\newcommand{\Tmax}{T_{\textrm{max}}}
\newcommand{\Kpw}{K_{\textrm{pw}}}

\renewcommand{\thesection}{\Alph{section}}

\begin{document}

\preprint{APS/123-QED}

\title{A coupled-oscillator model for the formation of planetary rings}

\author{Ruoming Gong}
\email[E-mail me at: ]{ruoming.gong@northwestern.edu}
\affiliation{Department of Engineering Sciences and Applied Mathematics, Northwestern University, Evanston, IL, USA}
\author{Theodore Broeren}
\affiliation{Non-Kinetic Capabilities \& Technologies, RTX BBN, Arlington, VA, USA}
\author{Eryn M.~Cangi}
\affiliation{Laboratory for Atmospheric and Space Physics, University of Colorado, Boulder, CO, USA}
\author{Daniel M.~Abrams} 
\affiliation{Department of Engineering Sciences and Applied Mathematics, Northwestern University, Evanston, IL, USA}
\affiliation{Northwestern Institute for Complex Systems, Northwestern University, Evanston, IL, USA}
\affiliation{Department of Physics and Astronomy, Northwestern University, Evanston, IL, USA}
\email[E-mail me at: ]{dmabrams@northwestern.edu}

\begin{abstract}
We study the dichotomy between compact satellite and ring formation in proto-planetary disks.  Specifically, we examine the behavior of a model system of $N$ identical particles locked into circular, gravitationally-bound orbits around a central body. We treat interactions as dominated by inter-particle collisions, and extract an effective two-particle interaction function based on both theory and simulations. We then demonstrate that the expected dynamics are equivalent to a variant of the Kuramoto model, which undergoes a phase transition as parameters vary.  This offers a novel potential explanation for the transition between formation of rings versus moons. 
\end{abstract}

\maketitle

\section{Introduction}
\begin{figure}[t!] 
  \centering
  \begin{tabular}{c c c}
  \includegraphics[width=0.315\columnwidth, trim=95 50 95 50, clip]{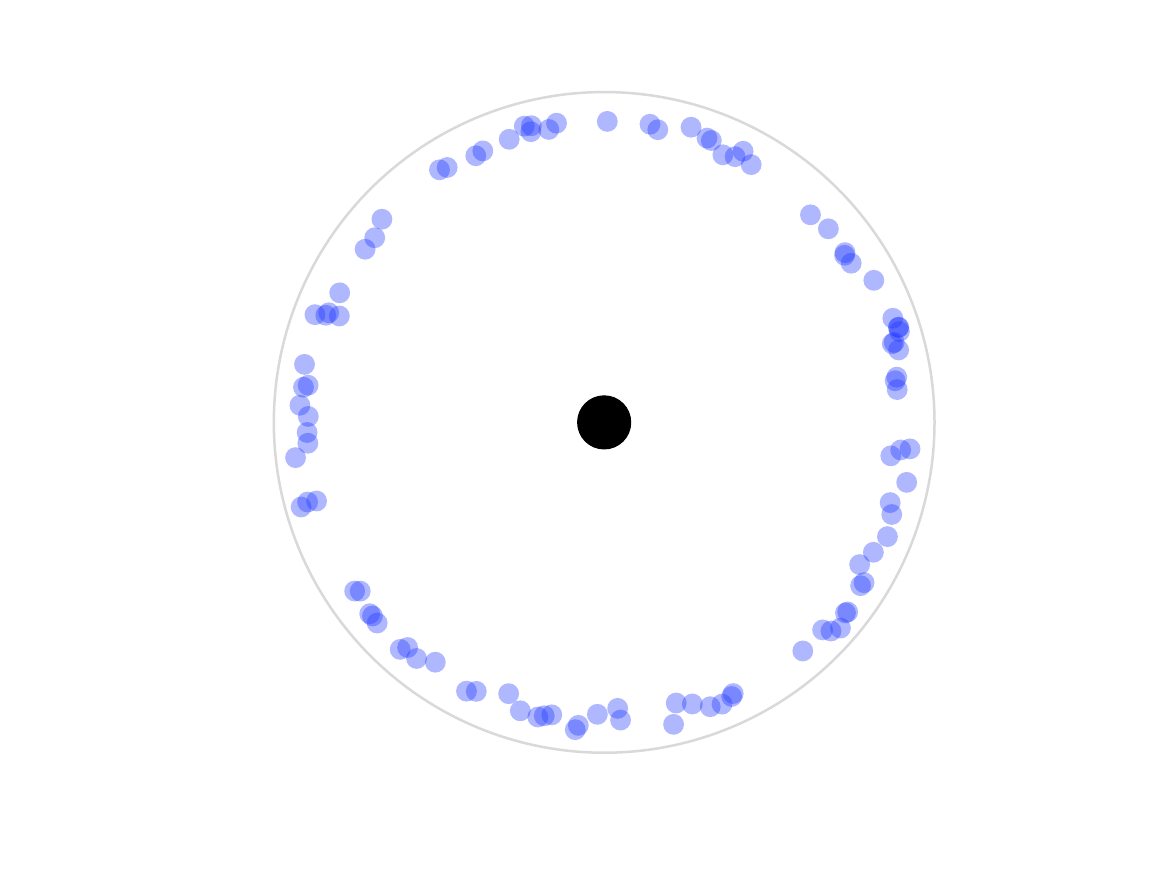} & \includegraphics[width=0.315\columnwidth, trim=95 50 95 50, clip]{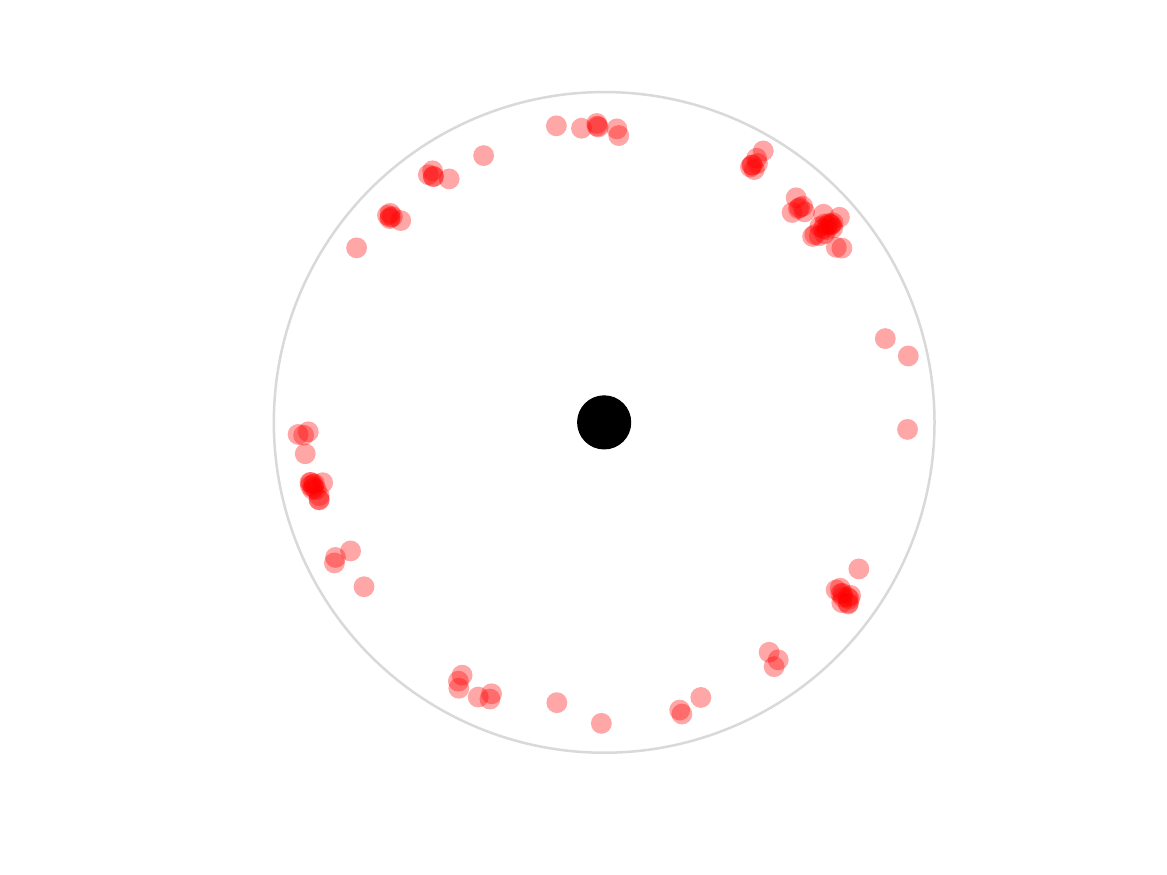} & \includegraphics[width=0.315\columnwidth, trim=95 50 95 50, clip]{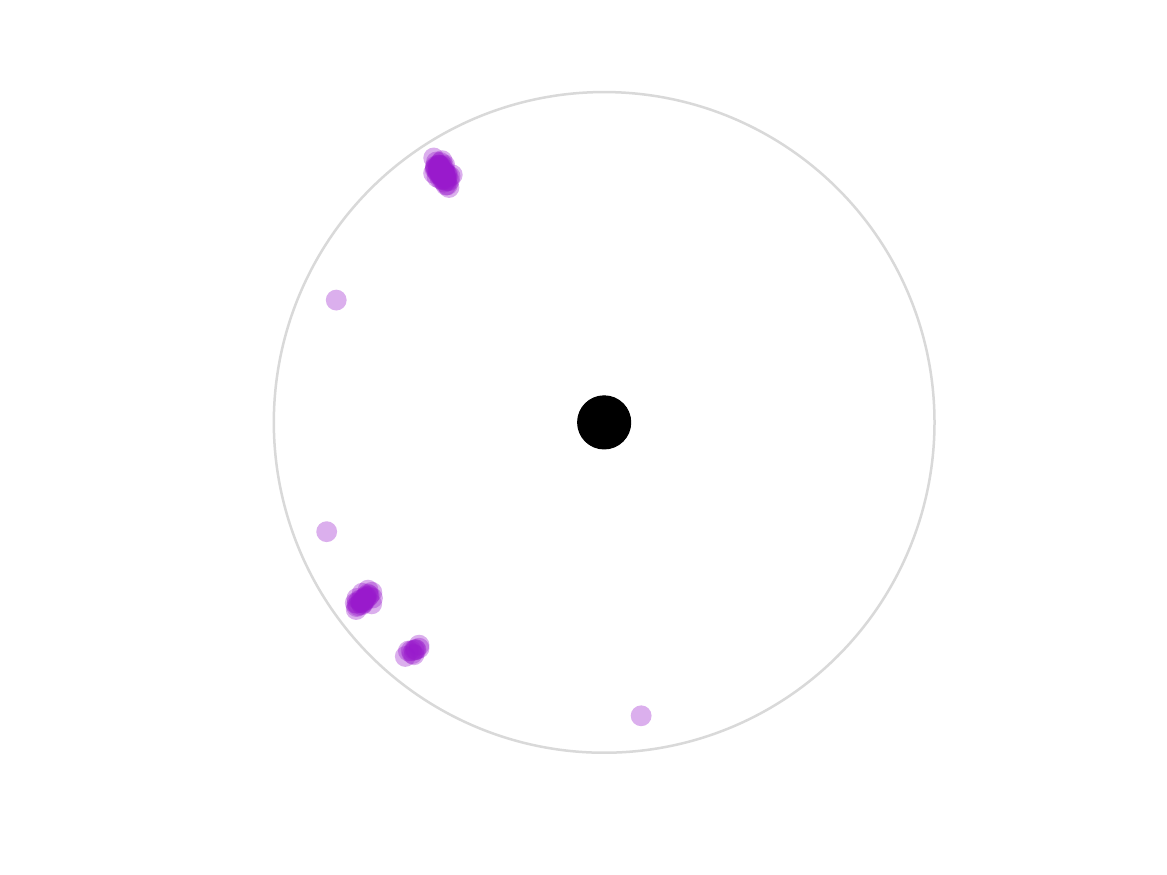}
  \end{tabular}
  \caption{\textbf{Typical simulation results}. A distribution of \textit{N} particles with uniformly randomly distributed initial angles and normally distributed initial radii was simulated as described in section \ref{sec:sim}. All particles moved in counterclockwise circular orbits with initial angular velocities $\omega_i = \sqrt{G M / r^3}$. Panels show initial conditions, a snapshot after 12,000 collisions with order 0.75, and the steady state of the system.  For these simulations $N=100$, $d=10^6$ m, $\mu_r=10^9$ m, $\sigma_r=2 \times 10^7$ m, and the mass of the central body is $M = 5.683\times10^26$ kg. }
  \label{fig:example}
\end{figure}

Saturn's large ring system has been a subject of intense scientific interest for hundreds of years (for brevity we cite some reviews of the extensive literature: \cite{Goldreich1982, Orton2009, Charnoz2009, Esposito2010, Spilker2019}). A puzzle remains, however, as to why Saturn has such a robust system of rings, while other gas giants have thinner and fainter rings, and the rocky planets have no significant rings at all \cite{Burns1984, dePater2006}. Two prominent theories involve either an ancient moon getting destroyed and spewing its mass around the planet before it spiraled into Saturn \cite{Canup2010}, or passing comets becoming captured by the planet's gravity and eventually breaking up \cite{Dones1991}. More recent research suggests the ring may be formed from the debris generated by the collisions of the first generation moons accreted from the rings in previous theories \cite{Ida2019, Teodoro2023}. A major factor in our uncertainty regarding ring systems is our limited ability to estimate their ages \cite{Esposito1986, Iess2019, Cuzzi2023}. Are rings we see today a transient phenomenon, or will they stably persist for an indefinite amount of time? Ultimately, we would like to understand the conditions under which mass in orbit will form a ring system like that shown in the left panel of Fig.~\ref{fig:example}, as opposed to a system of moons like that shown in the right panel of Fig.~\ref{fig:example}.

One key consideration in differentiating where rings and moons can form is the Roche limit \cite{Esposito1993}, which defines a boundary within which tidal forces hinder the aggregation of small particles into large moons. 

Until recently all dense rings were thought to lie within the Roche limit, however, recent observations may violate that constraint \cite{Morgado2023}.
This suggests new ideas are needed to elucidate the formation mechanisms of both rings and moons. 

Many previous efforts to decipher the origin and evolution of Saturn's rings have employed fluid approximations \cite{Cuzzi2010, Salmon2010}. These approaches do not attempt to model the collisions of the ring particles explicitly. A complementary line of work has used local and global $N$-body simulations, as well as particle-based granular models, to study dense ring dynamics, self-gravity wakes, viscosity, and viscous overstability at the particle level \cite{Rein2013, Ballouz2017, Lehmann2024, Salo2025}. While some studies have incorporated collisions into their simulations (e.g., \cite{Hyodo2017, Dubinski2019}), they have not conducted in-depth mathematical analysis of these interactions. In contrast, our approach explicitly addresses the collisions and focuses on an analytically tractable model that is fundamentally based on these collisional dynamics.

Rather than attempt to faithfully reproduce the detailed physics of the system, we retain only what we consider key physical aspects of the dynamical phenomenon.  Specifically, we treat a circumplanetary disk as made up of a finite set of interacting particles restricted to circular orbits.  The angular momentum of the system remains conserved while energy may be lost through dissipative collisions.  This energy change dynamics can lead to a nonuniform angular distribution of matter in space under certain conditions.  We compare simulations of this model to an even simpler abstracted system of coupled phase oscillators---the Kuramoto model---to gain analytical insight. 

The Kuramoto model (KM) is a paradigmatic mathematical model for synchronization of coupled oscillators \cite{Kuramoto1975, Strogatz2000, Acebron2005}. When defined on a network, simple phase oscillators with natural frequencies drawn from a known distribution $g(\omega)$ are coupled together and interact continuously via a sinusoidal coupling kernel:
\begin{equation} \label{eq:KM}
    \frac{d\theta_i}{dt} = \omega_i + \frac{K}{N}\sum_{j=1}^N A_{ij}\sin(\theta_j-\theta_i). 
\end{equation}
Here oscillator $i$ with angle $\theta_i(t)$ has fixed natural frequency $\omega_i$, and is coupled to its neighbors (defined by adjacency matrix $\mathbf{A}$) with coupling strength $K$. We compare the KM with our ring evolution model by imposing the same interaction network structure and estimate the effective coupling strength from simulation.

\section{Collision-based model}

For simplicity, we begin by assuming that particles' orbits about a central body of mass $M$ are perfect coplanar circles.  Thus particle speed $v = \omega r$, where $\omega$ is the particle's orbital angular velocity and $r$ its orbital radius. Then angular momentum $L = I\omega = m r^2 \omega = r m v$ for a single particle of mass $m$.
For a circular orbit centripetal forces are exactly balanced by gravitational forces, so $mv^2 / r = GMm / r^2$; thus $v = \sqrt{GM / r}$, and $L = m\sqrt{GMr}$.

Previous research has shown that mass in a ring system generally moves inward while angular momentum moves outward, assuming conservation of momentum and energy \cite{Lynden-Bell1974}. Motivated by that observation, when two particles of masses $m_1$ and $m_2$ collide, we model the collision as exactly conserving angular momentum, but with a small fraction of mass ejected to a large distance, thus carrying angular momentum outward. That is, for momentum, we impose  $L_{\textrm{initial}} = L_{\textrm{final}}$, i.e., 
\begin{align}
	&m_1 \sqrt{\rione} + m_2 \sqrt{\ritwo} \nonumber\\
    = &\left(m_1-\frac{dm}{2}\right) \sqrt{\rfone} + \left(m_2-\frac{dm}{2}\right) \sqrt {\rftwo}+ dm\sqrt{r_3},
    \label{eq:momentum}
\end{align}
where $\rione$  and $\rfone$ represent the orbital radius of particle 1 before and after the collision, respectively, and $dm$ is the small mass ejected to a large distance $r_3 \gg r_{1,2}$.  In simulation we track only the particles not ejected to this large distance.

The energy of each particle is the sum of its gravitational potential and kinetic energies, $E_{\textrm{potential}} + E_{\textrm{kinetic}} = -G M m / r + \frac{1}{2} m v^2$. Here we ignore all other sources of energetic changes including, e.g., particle spin, chemical, nuclear, and thermal effects.  Rewriting kinetic energy in terms of radius, we thus have 
\[	
    E_{\textrm{total}} =  -\frac{GMm}{r} + \frac{GMm}{2r} = -\frac{GMm}{2r}
\]
(also a consequence of the virial theorem \cite{clausius1870ueber}).

Energy balance then implies 
\begin{align} \label{eq:energy}
	&-\left(\frac{m_1}{\rione} + \frac{m_2}{\ritwo}\right)(1+\epsilon') = \nonumber \\
     &\qquad-\left(\frac{m_1-dm/2}{\rfone} + \frac{m_2-dm/2}{\rftwo} + \frac{dm}{r_3}\right) ,
\end{align}
where we allow for a small amount of dissipation through the parameter $\epsilon' > 0$, taking $|\epsilon'| \ll 1$, and we define the fraction of energy possessed by the ejected small mass $dm$ as 
\begin{equation}
    \epsilon = \frac{dm/r_3} {m_1/\rione + m_2/\ritwo}.
    \label{eq: def_eps}
\end{equation}

Solving Eqs.~\eqref{eq:momentum} and \eqref{eq:energy} while restricting to $0 \leq \epsilon \ll 1$ and $m_1, m_2,\rione, \ritwo, \rfone, \rftwo > 0$, we find the post-collision radius $\rfone$ to be the square of the second smallest positive real root of a quartic polynomial, taking $\rione < \ritwo$ without lose of generality.
Exact coefficients of that polynomial are given in terms of problem parameters in Supplemental Material (SM) section \ref{SM:exact_polynomial}. 
%
%
%
Approximate solutions when $\epsilon'=0$ (no energy dissipation) are: 
\begin{align} \label{eq:approxsol}
\begin{split}
  \rfone &= r_{1} +  r_1 \frac{\sqrt{r_{3}}}{r_{2}^{3/2} - r_{1}^{3/2}} \left( \frac{m_2}{m_1} r_1 + r_2 \right) \frac{r_3}{r_2}\epsilon + \mathcal{O}(\epsilon^2), \\
  \rftwo &= r_{2} - r_2 \frac{\sqrt{r_{3}}}{r_{2}^{3/2} - r_{1}^{3/2}} \left( r_1 + \frac{m_1}{m_2} r_2 \right) \frac{r_3}{r_1}\epsilon + \mathcal{O}(\epsilon^2),
\end{split}
\end{align}
(where we drop superscripts on initial radii for clarity and assume $r_1 \neq r_2$).  When $\epsilon = 0$ (no ejected mass), approximation solutions are:
\begin{align} \label{eq:inelastic_approxsol}
\begin{split}
  \rfone &= r_{1} -  r_1 \frac{\sqrt{r_{2}}}{r_{2}^{3/2} - r_{1}^{3/2}} \left( \frac{m_2}{m_1} r_1 + r_2 \right) \epsilon' + \mathcal{O}(\epsilon'^2), \\
  \rftwo &= r_{2} + r_2 \frac{\sqrt{r_{1}}}{r_{2}^{3/2} - r_{1}^{3/2}} \left( r_1 + \frac{m_1}{m_2} r_2 \right) \epsilon' + \mathcal{O}(\epsilon'^2).
\end{split}
\end{align}

This shows that, without energy dissipation, orbits approach one another, and without ejected mass, orbits move apart.  When both energy loss and mass ejection occur simultaneously, the post-collision radii are determined by the relative strength of angular-momentum removal through ejected mass and energy loss through dissipation.

\section{Results} \label{sec:results}
\textbf{Simulation:} \label{sec:sim}
We numerically simulate $N$ particles in circular orbits about a central body, allowing for collisions only when their radial separation is less than a fixed value $d$ (a proxy for the particles' effective diameters).  Initial conditions for the simulation are uniformly random initial angles $\theta_i$, identical masses, and normally distributed initial radii $r_i \sim \mathcal{N}(\mu_r, \sigma_r)$. 

In order to advance the simulation through time, we use an event-driven approach, ``jumping'' from collision to collision and skipping the time in between. This is done by calculating the time until next collision $\Delta t_{ij}$ for all pairs of particles $i,j$ that orbit sufficiently closely to interact ($|r_j-r_i| \leq d$), then selecting the smallest $\Delta t_{ij}$ (i.e., the minimum positive value of $\{(\theta_i - \theta_j + \phi)/(\omega_j - \omega_i): \phi \in \{-2\pi,0,2\pi\}\}$). Time is advanced to this point, so $\theta_i^{\textrm{new}} = \theta_i^{\textrm{old}} + \omega_i \Delta t ~ \forall ~ i$. The two particles involved in the collision are instantaneously shifted to new radii computed via Eqns.~\eqref{eq:momentum} and \eqref{eq:energy}, where $\epsilon = \chi \epsmax$. Here $\epsmax$ is the maximum fraction of energy that can be carried by the ejected particle while satisfying the conservation equations (i.e., preserving real roots). We choose $0 < \chi < 1$ uniformly at random at each step.  

If two particles repeatedly interact and approach each other in radius (e.g., when $\epsilon'=0$), the average amounts that the radii change per collision decrease (note that $\epsmax \to 0$ as $r_1 \to r_2$). It is natural to interpret this as a scenario where particle aggregation occurs (though our results hold even when aggregation in not allowed, see SM sec.~\ref{sec:withandwithout_aggregation}). Our aggregation scheme groups particles $i$ and $j$ together if they collide and $|r_i-r_j| < \delta r$, where the value of $\delta r$ is chosen to be small enough so $\delta r \ll d$. Grouping of particles $i$ and $j$ entails replacing both with a single particle $k$ of mass $m_k = m_i + m_j$ and radius $r_k = (m_i r_i+m_j r_j)/(m_i+m_j)$ (also $\theta_k = \theta_i = \theta_j$ since particles align at the time of collision/aggregation).

We also incorporate energy dissipation and particle disaggregation into the simulations as a consequence of nonconservative collisions. We assume energy is dissipated at a constant rate $E_0$, so the energy loss at each event is $E_0 \Delta t$, where $\Delta t$ is the time between consecutive events. At each event time, we choose one particle randomly and treat it is a result of collision between two equal-mass particles initially located at the same radius; Eqns.~\eqref{eq:momentum} and \eqref{eq:energy} give the new radii of two resulting daughter particles, with $\epsilon'$ chosen such that exactly $E_0 \Delta t$ energy is lost.  In practice, when too many small particles accumulate the simulation can become unmanageably slow, so when a randomly selected particle is below a threshold $m_{\textrm{min}}$, instead of disaggregation we choose a second particle within its radial interaction range ($|r_j-r_i| \leq d$) and update both particles' orbits according to Eqns.~\eqref{eq:momentum} and \eqref{eq:energy}(again so that exactly $E_0 \Delta t$ energy is lost).

We emphasize that here disaggregation is associated with energy loss---not gain.  This is because we consider only orbital energy in this model. As stated earlier, we ignore all other sources of energetic changes (e.g., particle spin, deformation, chemical, nuclear, and thermal effects).

We also note that, in this model, particles only interact in a pairwise fashion at discrete points in time (collision events)---not continuously. Therefore, in order to compare to the Kuramoto model, where all particles are continuously interacting with all others, we treat the cumulative effect of many instantaneous collisions as mimicking a continuously operating interaction function. Numerically, we infer this function by running thousands of short simulations with varying initial conditions. 

\begin{figure}
    \centering
    \includegraphics[width=0.8\columnwidth]{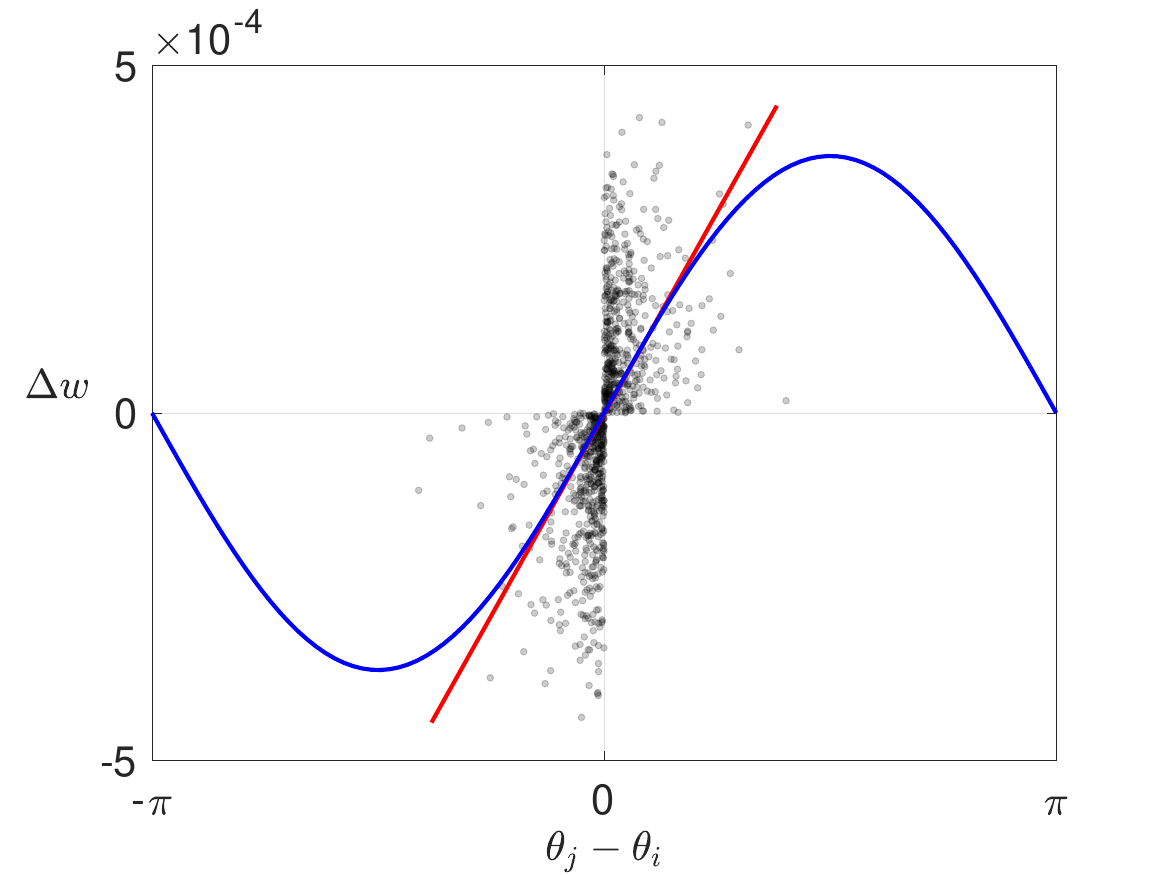}
    \caption{\textbf{Effective pairwise interaction among particles}. We simulate the collision model and then fit a sinusoidal function as a simplified tractable model. The fit is done by regression on the slope at the origin.  Parameters for simulation: $N = 100$, $\mu_r = 1$, $d = 0.0004$, and $\sigma_r = 0.02$ ($d / \sigma_r = 0.02$).}
    \label{fig:couplingstrength}
\end{figure}

In each short simulation, we initialize the system with a uniform random angular distribution and Gaussian random radial distribution of particles as described above. We then compute the angular difference, $\Delta\theta = \theta_2(0)-\theta_1(0)$, and the change in angular frequency, $\Delta\omega = \omega_j^i-\omega_j^f$, for the first pair of particles to collide. Here, $j$ is chosen so that $\Delta\theta$ and $\Delta\omega$ have the same sign. Fig.~\ref{fig:couplingstrength} shows the results of 100,000 such simulations (1,000 points sampled for visualization), along with the inferred effective mean interaction function, analogous to the sinusoidal coupling function in the Kuramoto model. 

\begin{figure}
    \centering
    \subfigure{\includegraphics[width=0.15\textwidth]{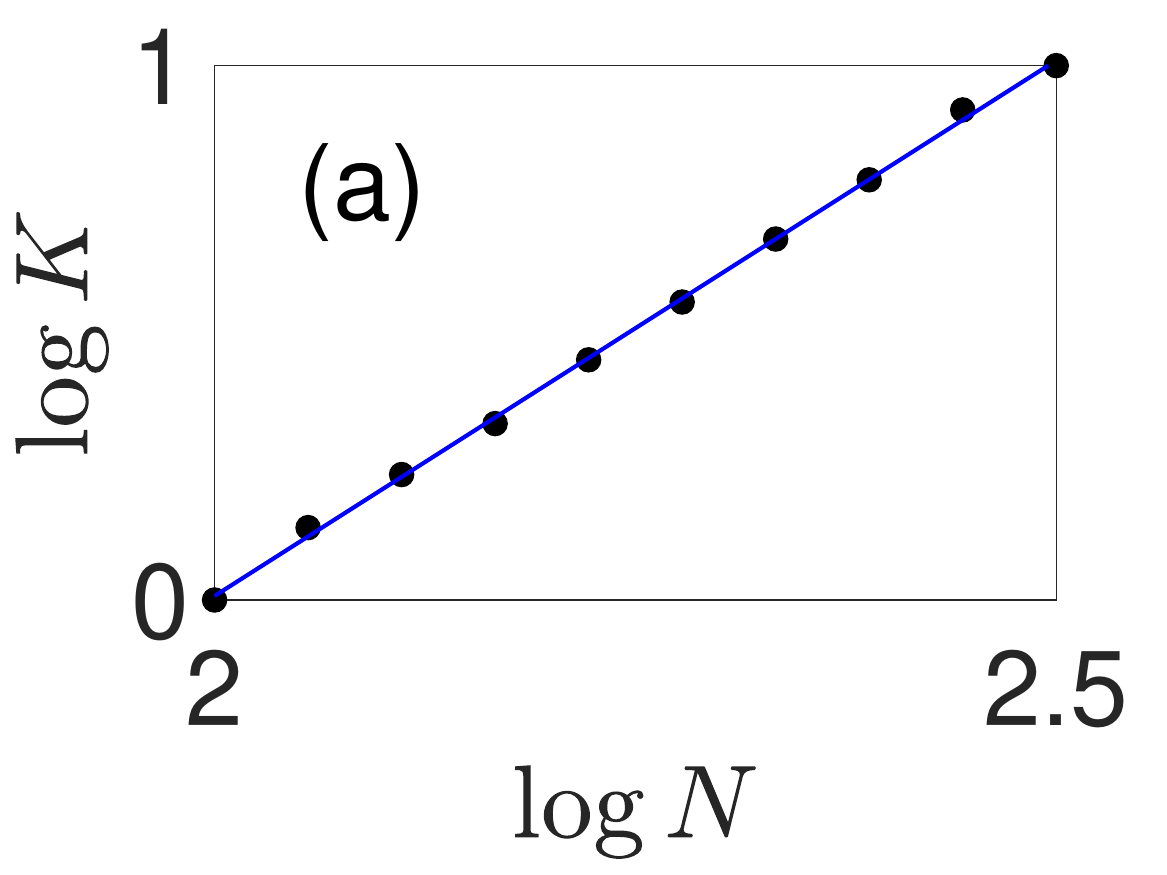}}
    \hspace{-3mm}
    \subfigure{\includegraphics[width=0.15\textwidth]{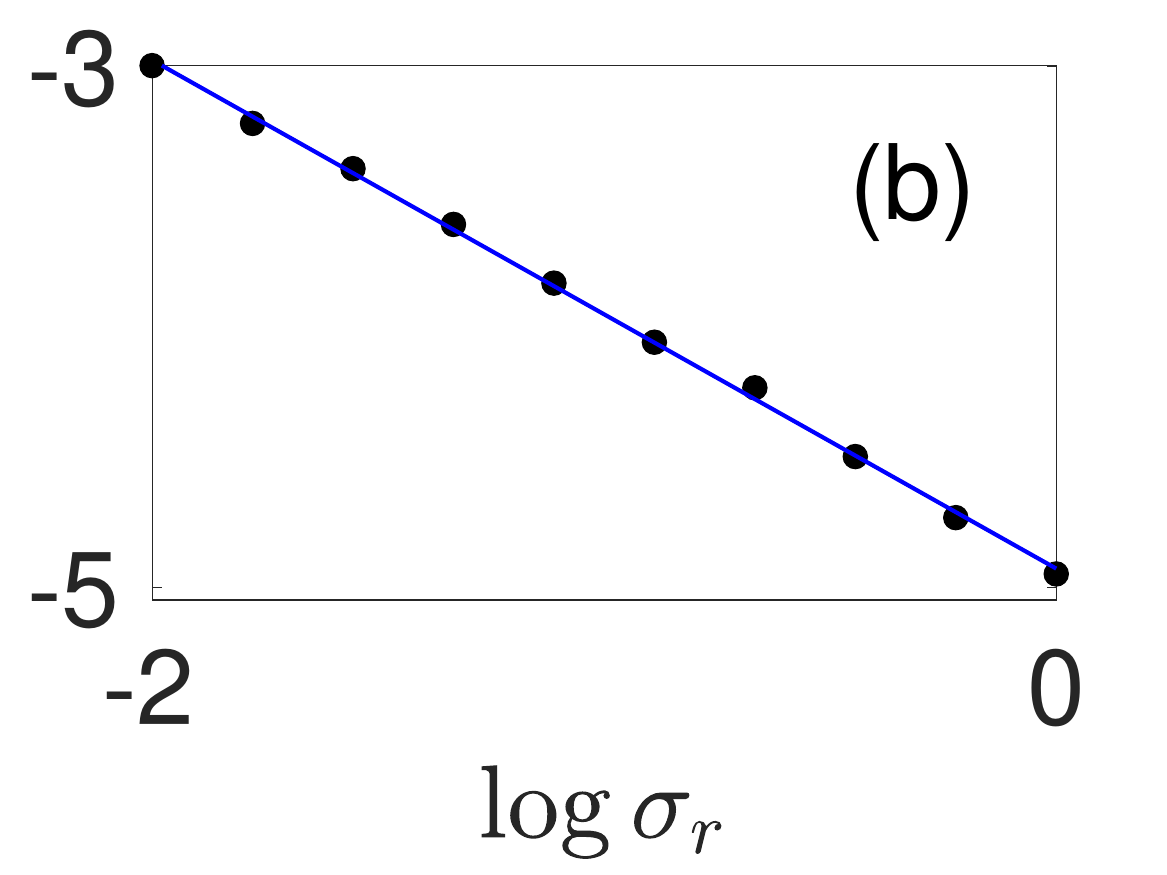}}
    \hspace{-3mm}
    \subfigure{\includegraphics[width=0.15\textwidth]{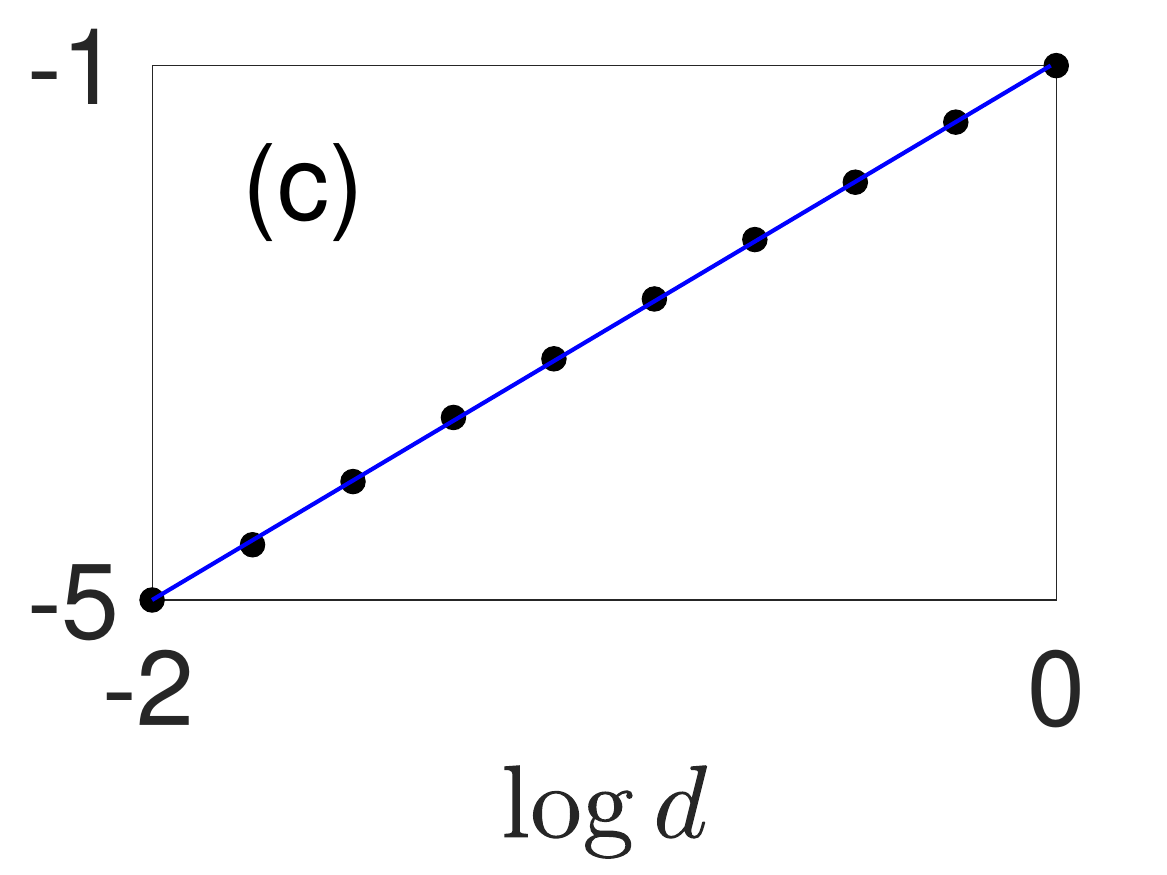}}
    \caption{\textbf{Relationship between effective pairwise coupling strength $\Kpw$ and simulation parameters.} (a) Effect of number of particles: $\Kpw \propto N^2$, (b) Effect of std.~dev.~of particle distribution:  $\Kpw \propto \sigma_r^{-1}$, (c) Effect of interaction range: $\Kpw \propto d^2$. }
    \label{fig:scaling_argument}
\end{figure}

\textbf{Coupling strength:}\label{sec:coupling_strength}
The fitted sine wave in Fig.~\ref{fig:couplingstrength} changes with number of particles $N$, interaction distance $d$, and with the mean $\mu_r$ and standard deviation $\sigma_r$ of the particle radial distribution.  Quantifying the dependence of the slope at the origin (which we refer to as $\Kpw$---the pairwise coupling strength) on each of these parameters yields
\begin{equation}  \label{eq:Keff}
  \Kpw \propto \frac{N^2d^2}{\sigma_r\mu_r}.
\end{equation}
This empirical scaling (see Fig.~\ref{fig:scaling_argument}) can also be justified on a theoretical basis.  We present detailed arguments in SM sec.~\ref{sec:2D_scalingargument}, but the brief idea is that the coupling strength is set by how frequently collisions occur.
For a given particle, the expected time between collisions,
$\mathbb{E}(T_{\mathrm{coll}})$, is approximately $T_{\mathrm{orb}}/\tau$, where
$T_{\mathrm{orb}} = 2\pi/\omega$ is the particle's orbital period and $\tau$ is
the system's optical depth,
\begin{equation}
    \tau \;=\; \frac{(\text{number of particles})(\text{area of each particle})}
    {\text{area covered by the ring}}.
\end{equation}

\textbf{Discrete Kuramoto model:}
A natural coupled oscillator analog of the collision model is the discrete Kuramoto model with pairwise interaction, which is given (for the fully connected network) by
\begin{equation}\label{eq:pairwiseKM}
    \Delta \theta_i = \omega_i  \Delta t + \Kpw \sum_{j=1}^N \delta_{ik}\delta_{jl}\sin(\theta_j - \theta_i)  \Delta t\,, 
\end{equation}
$i=1\ldots N$.
At each time step we randomly choose a pair of interacting oscillators $(k,l)$; these two influence one another according to sinusoidal coupling while all other oscillators evolve only based on their natural frequency.  The pairwise KM has similar dynamics to the KM up to a rescaling of $\Kpw$, with $K$ in the original KM scaling as $\Kpw / N$ (see SM sec.~\ref{sec:pariwise_KM} for details). 

With all-to-all coupling, the original KM has exact solution in the $N \to \infty$ limit:
\begin{equation}\label{eq:continuum condition}
    1 = K\int_{-\pi/2}^{\pi/2}\cos^2\theta g(K R \sin\theta)d\theta~,
\end{equation}
where $R \in [0,1]$ is the order parameter defined as $\left|N^{-1} \sum_j e^{i \theta_j}\right|$.
A synchronous branch of solutions bifurcates off the $R=0$ incoherent state at $K_c = 2 / \pi g(0)$ (evident here as $R \to 0^+$).  This phase transition represents a balance between the dispersive effects of heterogeneity (differing natural frequencies) and the synchronizing tendencies of coupling.

In an analogous way, the collision model also appears to show a phase transition when the synchronizing tendencies of collisions are balanced against the dispersive effects of differing orbital periods. However, in the collision model, unlike in the KM, angular velocities are not fixed.  We observe that lower heterogeneity in angular velocity (i.e., lower radial variance) correlates with higher order.

\begin{figure}[t!]
    \centering    
    \includegraphics[width=0.8\columnwidth]{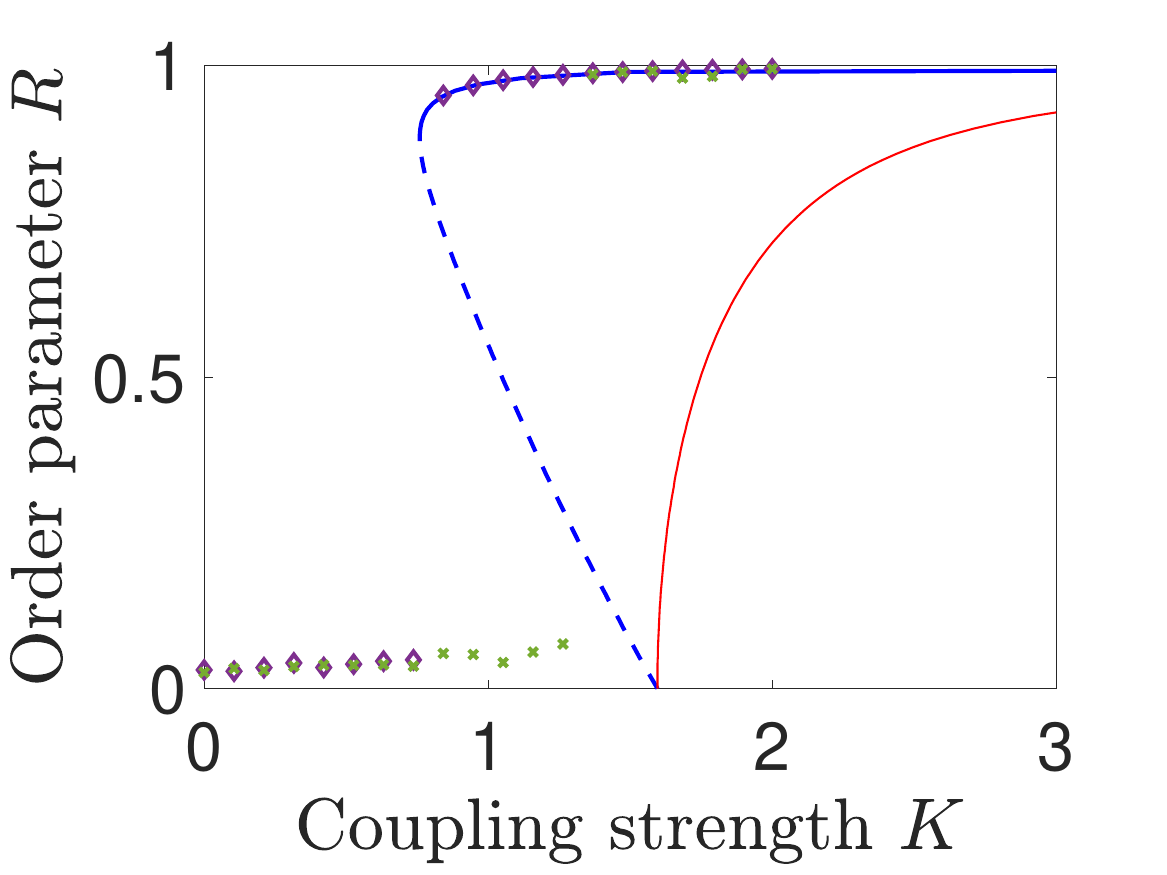}
    \caption{\textbf{KM simulation with natural frequency distribution dependent on order.} Natural frequency $\omega\sim\mathcal{N}(0,1-\alpha r)$, $\alpha=0.8$.  The blue curve is the theoretical prediction with $\alpha = 0.8$ (solid=stable, dashed=unstable). The red curve is the theory for standard KM. Purple diamonds from numerical simulation with decreasing K, green crosses from numerical simulation with increasing K ($N=1000$). }
    \label{fig:backward_simulation}
\end{figure}

\begin{figure}[t!]  
  \centering
  \includegraphics[width=0.8\columnwidth]{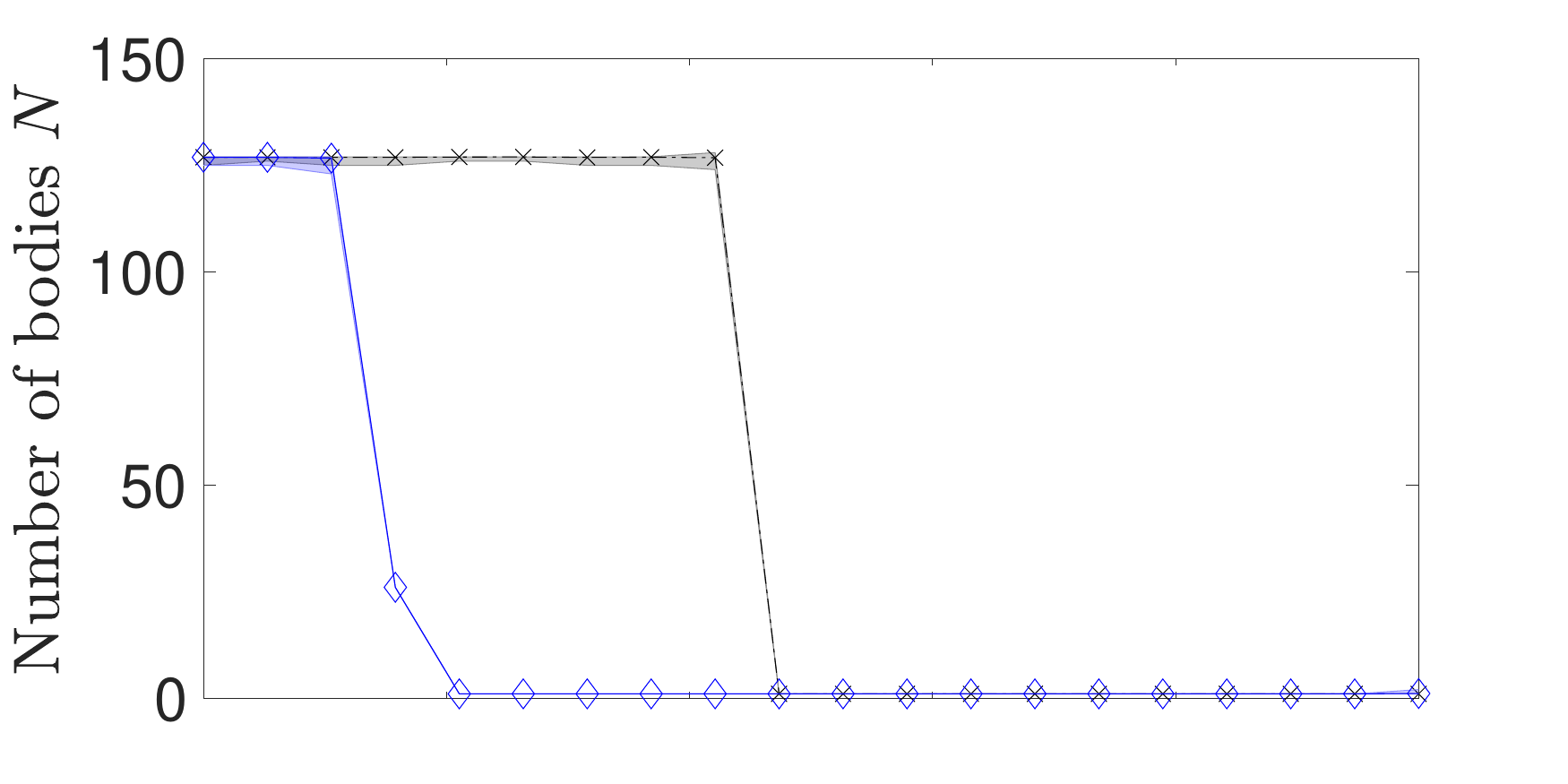} 
  \includegraphics[width=0.8\columnwidth]{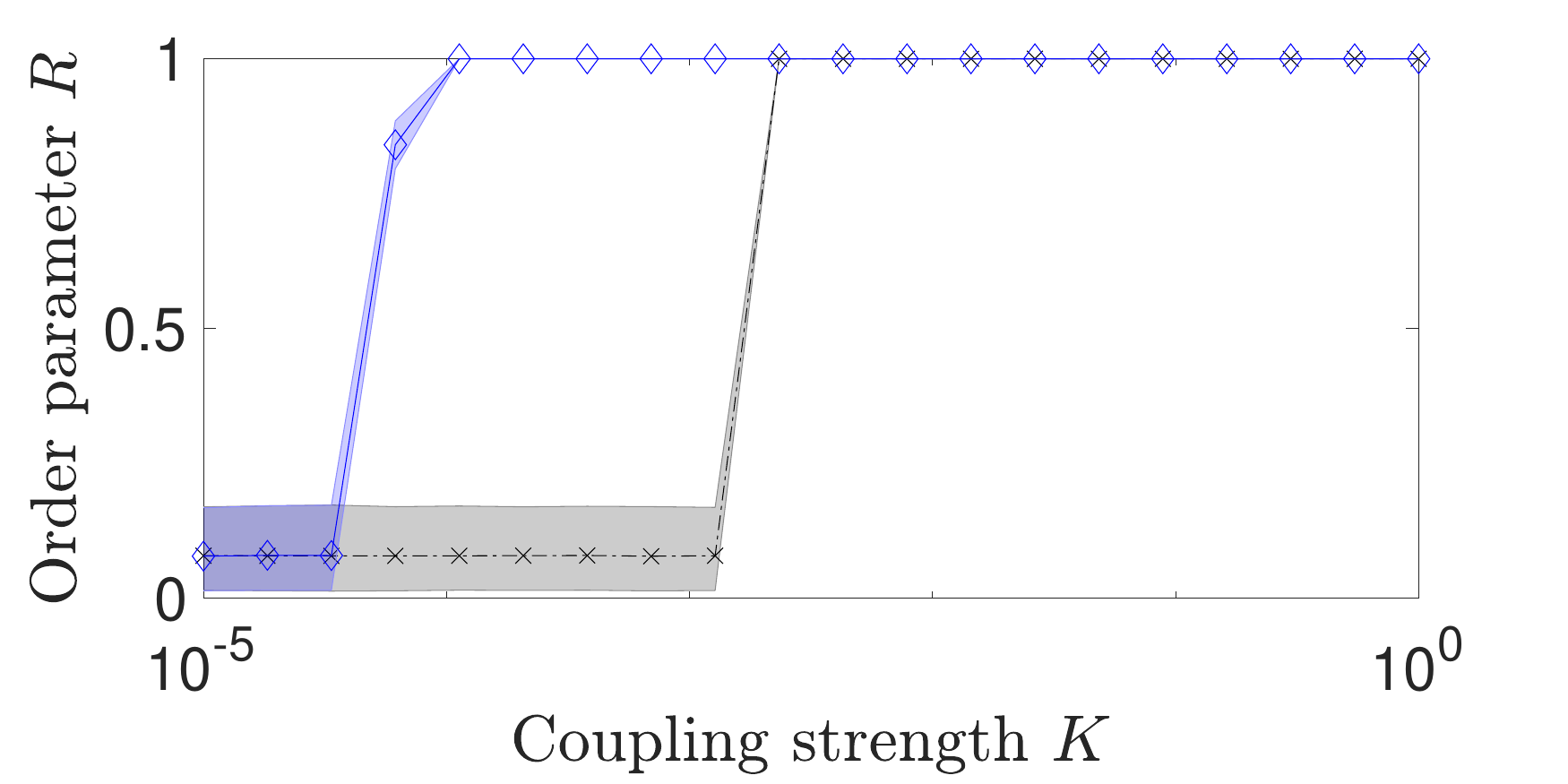} 
  \caption{\textbf{Phase transition in collision model.} Steady-state order for the particle orbit simulation with energy injection (blue squares). $ d/\sigma_r = 2, \Tmax = 10$. $K$ is varied by changing $\mu_r$ where $K\propto\mu_r^{-5/2}$. Blue curve is plotted with decreasing $K$ and black curve is plotted with increasing $K$. 
  }
  \label{fig:hysteresis_energy}
\end{figure}

To incorporate such an effect into the KM, we consider the case where $g(\omega)$ is normally distributed and take the variance to decrease linearly as $R$ increases, i.e., $g(\omega) = \exp{(-\omega^2/2\sigma_\omega^2)}/\sqrt{2\pi\sigma_\omega^2}$ with $\sigma_\omega = \sigma_\omega(R) = \sigma_{\omega,0}(1-\alpha R)$. We solve Eq.~\eqref{eq:continuum condition} numerically under that assumption; Fig.~\ref{fig:backward_simulation} shows that it implies bistability of the system: for  $K<K_c$ it is possible to have either a synchronous or an incoherent state (analogous to moon or ring states) depending on the initial condition of the system. 
Fig.~\ref{fig:hysteresis_energy} shows that a similar bistability can occur in the collision model.

We can also view the competing effects leading to the phase transition through the lens of energy.  The dispersion associated with variation in orbital period is mitigated when masses move to closer orbits ($|r^f_2-r^f_1| < |r^i_2-r^i_1|$), which occurs as energy is effectively added to the system (focusing on the objects near the central body) through the ejection of small masses to a far orbit.  However, as energy is lost through dissipative collisions, the opposite happens ($|r^f_2-r^f_1| > |r^i_2-r^i_1|$), leading to a possible balance between dispersive and synchronizing tendencies.

For the collision model in a fully connected system (all bodies can collide with all others) and no energy dissipation, the unique equilibrium state is a single aggregated body. This is analogous to the Kuramoto model with identical oscillators, where any nonzero coupling leads to perfect synchrony. To avoid aggregation into a single moon, energy loss must occur. Although energy is not defined for the KM, fluctuations around the equilibrium state in the discrete Kuramoto model mimic the balance between aggregation and dissipation in the collision model.

\textbf{Phase transition:}
Similarly to the pairwise KM, we also observe a phase transition in the collision model. In Fig.~\ref{fig:hysteresis_energy} we plot both the number of bodies $N$ and the \textit{final order} $R$ of the particle simulation for the collision model. The \textit{final order} is defined as the time averaged magnitude of the order parameter $\sum_j m_j e^{i\theta_j} / M_{\textrm{tot}}$ (i.e., the physical center of mass), averaged over time after collisions have ceased to occur\footnote{Note that we cannot simply use the final value of order for the particle simulation because that value oscillates in time as the groups of particles rotate around the central body.}. The effective coupling strength $K$ is modified by varying $\mu_r$ so that the initial network structure remains unchanged. The hysteresis loop in Fig.~\ref{fig:hysteresis_energy} seems to indicate a first-order phase transition, consistent with the theoretical prediction of the pairwise Kuramoto model (with apparently $\alpha$ near 1). This supports the analogy between the pairwise KM with order-dependent heterogeneity and the collision model.

\section{Discussion and conclusions}
We have introduced a reduced collision-based framework for studying the transition between ring-like and moon-like outcomes in orbiting particulate systems. The central idea is that repeated pairwise collisions can act as an effective synchronizing interaction: collisions tend to reduce differences in orbital radius and angular velocity, while dissipative processes tend to spread particles apart. The competition between these two tendencies produces behavior analogous to the synchronization transition in the Kuramoto model. 

This perspective differs from many traditional treatments of planetary rings, which often rely on fluid or continuum descriptions, and our results should not be interpreted as a complete theory of planetary ring formation. The model makes strong simplifying assumptions. For example, particles are confined to coplanar circular orbits, energy loss at a constant rate, etc. The Kuramoto model comparison is therefore an analogy, not an exact reduction from celestial mechanics. Nevertheless, the analogy is useful because it identifies a possible organizing principle: the ring-moon transition can be viewed as a balance between collision-induced synchronization and dispersive processes.

This viewpoint is complementary to, rather than a replacement for, the Roche-limit picture. The Roche limit gives a fundamental criterion for whether tidal forces suppress the growth of large aggregates, and it remains central to understanding why dense rings are commonly found close to their host planets. However, recent observations of dense rings outside classical Roche limits suggest that additional mechanisms may influence whether material remains ring-like or accretes into satellites. Our results suggest that collisional dynamics and the effective interaction network among particles may provide one such mechanism. In this sense, the present model may help explain why the same broad ingredients (orbiting particulate matter, dissipative processes) can lead to qualitatively different outcomes in different parameter regimes.

There are also natural connections to other orbital systems. Co-orbital moon systems such as Janus and Epimetheus (see, e.g., \cite{tiscareno2009rotation}) illustrate that bodies in nearby orbits can undergo long-lived, nontrivial dynamical interactions without simply merging. Extending the model to include gravitational exchanges, resonant interactions, and finite-size effects could help determine when nearby orbiting bodies merge, remain separated, or maintain recurrent exchange dynamics.

Finally, although our model is motivated by planetary ring and moon formation, the same conceptual framework may be relevant at other scales. More broadly, one may ask whether the competition between interaction-driven aggregation and dispersive processes also helps organize structure formation in other gravitational systems, such as accretion disks, solar systems, or even galactic-scale systems. We do not claim that the present model applies directly to these settings, but the analogy suggests a possible direction for future work.

\begin{acknowledgments}
\textit{Acknowledgements}---The authors would like to thank Yoram Lithwick for useful discussions and valuable feedback on the modeling approach, and Jennifer Ruda for her contributions to the numerical simulations.  We also acknowledge NSF support for E.M.C. through grant AST-1359462. 
\end{acknowledgments}

\bibliographystyle{apsrev4-2}
\bibliography{astrorefs.bib}

\clearpage
\newpage
\onecolumngrid
\suppressfloats

\setcounter{section}{0}
\setcounter{figure}{0} 
\setcounter{equation}{0} 
\renewcommand{\thefigure}{S\arabic{figure}} 
\renewcommand{\thesection}{S\arabic{section}} %
\renewcommand{\theequation}{S\arabic{section}.\arabic{equation}} 

\makeatletter
\renewcommand{\section}{\@startsection{section}{1}{\z@}%
    {-3.5ex \@plus -1ex \@minus -.2ex}
    {2.3ex \@plus .2ex}
    {\normalfont\large\bfseries\sffamily\lowercase}} 
\makeatother

\makeatletter
\renewcommand{\section}{\@startsection{section}{1}{\z@}%
    {-3.5ex \@plus -1ex \@minus -.2ex}
    {2.3ex \@plus .2ex}
    {\normalfont\large\bfseries}} 
\makeatother

\begin{center} 
    \Large{Supplemental Material:}

    A coupled-oscillator model for the formation of planetary rings
\end{center}

\section{Parallels to the pairwise Kuramoto Model}\label{sec:pariwise_KM}

In analogy with the collision model in the main text, here we define a pairwise variant of the Kuramoto model. We demonstrate how that pairwise model can be seen as equivalent to the standard KM with the correct choice of scaling of the coupling strength $\Kpw$.  

Before discussing the detailed relationship between $\Kpw$ and $K$ in the standard Kuramoto model, we first emphasize that $\Kpw$ depends explicitly on the number of oscillators $N$,
whereas $K$ is usually treated as a prescribed coupling parameter. This difference arises because $\Kpw$ is derived from a collision rate. Since coupling strength has units of inverse time, it is set by the frequency of interaction events. Increasing the number of particles reduces the typical inter-collision time for a given particle, thereby increasing the collision rate and producing a stronger effective coupling.


That said, a natural coupled oscillator analog of the collision model we present in the main text is the discrete Kuramoto model with pairwise interaction, which is given by (for the fully connected network)
\begin{equation}
    \Delta \theta_i = \omega_i  dt + \Kpw \sum_{j=1}^N \delta_{ik}\delta_{jl}\sin(\theta_j - \theta_i) dt~, 
\end{equation}
where at each time step we randomly choose a pair of interacting nodes $(k,l)$. These nodes interact according to the sine function while the phases of all other nodes are updated only based on their natural frequencies. 

This can also be expresed as
\begin{equation} \label{eq:better}
    \Delta \theta_i = \omega_i  dt + 
    \begin{cases}
        \Kpw \sin(\theta_1 - \theta_i) dt  & \text{with probability $2/N^2$}\\
        \Kpw \sin(\theta_2 - \theta_i) dt  & \text{with probability $2/N^2$}\\
        ...\\
        \Kpw \sin(\theta_N - \theta_i) dt  & \text{with probability $2/N^2$}\\
        0 & \text{with probability $1-(2N-2)/N^2$}
    \end{cases}~,
\end{equation}
or, with more compact notation,
\begin{equation} \label{eq:compact}
    \Delta \theta_i = \omega_i  dt + \frac{2\Kpw}{N} \left<\sin(\theta_j - \theta_i)\right>_j dt ~,
\end{equation}
where angle brackets indicate an average over all oscillators.\footnote{Technically we have approximated Eq.~\eqref{eq:better} in rewriting it as Eq.~\eqref{eq:compact}, though the approximation becomes exact for $N \to \infty$.}

\begin{figure}
    \centering
    \subfigure[]{\includegraphics[width=0.45\textwidth]{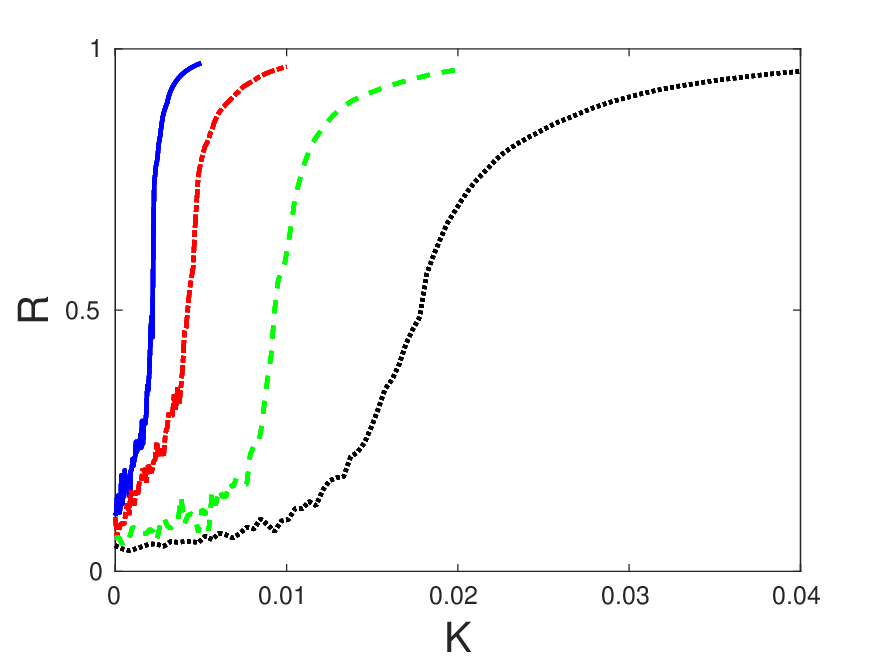}} 
    \subfigure[]{\includegraphics[width=0.45\textwidth]{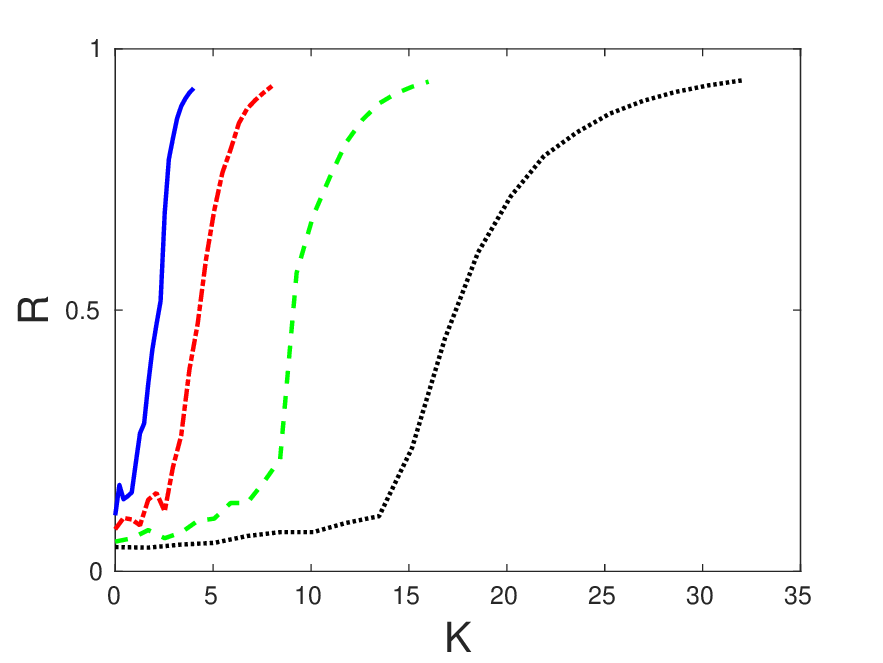}} 
    \caption{\textbf{KM with pairwise interaction at each step.} Here each data point 
    reflects a single simulation of the KM model with fixed K, using pairwise interaction at each simulation step.  Equilibrium order parameter is determined numerically. (a) Natural frequencies inferred from the collision model, with particle radii drawn from a normal distribution; simulations use Eq.~\eqref{eq:better}, (b) Natural frequencies drawn from a normal distribution; simulations use Eq.~\eqref{eq:better}. 
    }
    \label{fig:pairwiseCO}
\end{figure}

The critical $\Kpw$ in equation (\ref{eq:better}) scales linearly with $N$, as visible in Fig.~\ref{fig:pairwiseCO} (a)(b). We can compute the expected $\Delta \theta_i$ at each time step. Define $r,\psi$ such that $R e^{i\psi} = \left< e^{i \theta_j} \right>_j = \frac{1}{N}\sum_{j=1}^n e^{i \theta_j}$, then
\begin{align}
    \mathbf{E}(\Delta\theta_i) &= \omega_i dt + \frac{2}{N}\frac{\Kpw}{N}\sum_{j=1}^N\sin(\theta_j - \theta_i)dt \nonumber \\
    &= \omega_i dt + \frac{2}{N}\Kpw R \sin(\psi - \theta_i)dt .
\end{align}
Changing to a rotating frame defined by $\phi_i = \theta_i - \psi$, then we observe some oscillators locked (those with $|\omega_i - \bar{\omega}| \leq \frac{2}{N} \Kpw R$) and some drifting (those with $|\omega_i - \bar{\omega}| > \frac{2}{N} \Kpw R$).  The locked oscillators in the co-rotating frame will have fixed phases $\sin \phi_i^*  = \frac{N (\omega_i-\bar{\omega})}{2 \Kpw R}$. Ref.~\cite{Strogatz2000} shows that, for normally distributed $\omega_i \sim \mathcal{N}(0,\sigma)$, the critical $\Kpw = \sqrt{\frac{2}{\pi}}\sigma N$, consistent with the numerical simulations. 

To express Eq.~\eqref{eq:better} in terms of the standard KM model, we simply set $2\Kpw/N = K$, thus giving
\begin{equation}
    K = \frac{N}{2}\Kpw.
\end{equation}


\section{Coupling strength scaling argument}\label{sec:Couplingstrengthscalingargument}
In Sec.~\ref{sec:coupling_strength} we estimate the pairwise coupling strength $\Kpw$ as
\begin{equation}
    \Kpw = \frac{\mathbf{E}(\Delta \omega)}{\mathbf{E}(\Delta \theta)} .
\end{equation}
This follows from the assumption of a uniform phase distribution and is independent of the distribution of natural frequencies $\omega$. We now examine separately how $\mathbf{E}(\Delta \omega)$ and $\mathbf{E}(\Delta \theta)$ scale with the
model parameters.

\subsection{$\mathbf{E}(\Delta \omega)$ scaling}
Let $r_1, r_2$ be the orbital radii of two colliding particles, and define $\bar{r} = (r_1 + r_2)/2$ and $\Delta r = |r_2 - r_1| / 2$. Since collisions cause particles to move toward the mean orbital radius (and noting that $\omega \sim r^{-3/2}$), we have
\begin{equation}
    \Delta\omega \sim \bar{r}^{-3/2} - (\bar{r} + \Delta r)^{-3/2} \approx \frac{3}{2}\frac{\Delta r}{\bar{r}} \bar{r}^{-3/2}.
\end{equation}
And by rescaling time such that the orbital period is set to $1$, we obtain
\begin{equation}
    \Delta\omega \sim \frac{3}{2}\frac{\Delta r}{\bar{r}}.
\end{equation}
We have assumed $r$ is normally distributed with $r\sim\mathcal{N}(\mu_r, \sigma_r^2)$. So
\begin{equation}
    \mathbf{E}(\Delta \omega) \sim \frac{\mathbf{E}(\Delta r)}{\mathbf{E} (\bar{r})} = \frac{\mathbf{E}(\Delta r)}{\mu_r}
\end{equation}
and
\begin{equation}
    \mathbf{E}(\Delta r) = \frac{1}{2}\mathbf{E}\left(|r_1 - r_2| ~ \big| ~ |r_1 - r_2| < d \right),
\end{equation}
where $d$ is the interaction range. Note $r_1 - r_2 \sim \mathcal{N}(0, 2\sigma_r^2)$, so
\begin{align*}
    \mathbf{E}(\Delta r) &= \frac{2 \int_0^d x f_X(x) d x}{2 \int_0^d f_X(x) d x} 
    = \frac{\frac{2 \sigma_r}{\sqrt{\pi}}\left(1-e^{-d^2 /\left(4 \sigma_r^2\right)}\right)}{2 \Phi\left(\frac{d}{\sqrt{2} \sigma_r}\right)-1},
\end{align*}
where $f_X(x)=\frac{1}{2\sigma_r \sqrt{\pi}} e^{-x^2 / 4\sigma_r^2}$ and $\Phi$ is the cumulative distribution function of standard normal distribution.

In the weak interaction case, $d \ll \sigma_r$, then
\begin{equation}\label{eq:weak_interaction_omega}
    \mathbf{E}(\Delta r) \approx \frac{\frac{2 \sigma_r}{\sqrt{\pi}} \frac{d^2}{4 \sigma_r^2}}{\frac{d}{\sigma_r \sqrt{\pi}}}=\frac{d}{2}\sim d.
\end{equation}

In the strong interaction case, $d \gg \sigma_r$, then 
\begin{equation}\label{eq:strong_interaction_omega}
    \mathbf{E}(\Delta r) \approx \frac{\frac{2\sigma_r}{\sqrt{\pi}}}{1}=\frac{2\sigma_r}{\sqrt{\pi}}\sim \sigma_r.
\end{equation}

\subsection{$\mathbf{E}(\Delta \theta)$ scaling}
Suppose $\theta$ is uniformly distributed on $[0, 2\pi)$. Define $\Delta\theta = \min(|\theta_2-\theta_1|, 2\pi - |\theta_2-\theta_1|)$ which is the shorter angular distance on the circle. Then $\Delta\theta \sim \text{Unif}(0,\pi)$.

Now let $Q_i = \min_i(\Delta\theta_1,\Delta\theta_2,...,\Delta\theta_n)$, i.e., the $i^{\textrm{th}}$ smallest value in $\{\Delta\theta_1,\Delta\theta_2,...,\Delta\theta_n\}$. For $i = 1$, the minimum of this set, the CDF is given by
\begin{align*}
    F_{Q_1}(q) & = P(Q_1 \leq q) = 1 - P(Q_1 > q) \\ 
    & = 1 - P(\min(\Delta\theta_1,\Delta\theta_2,...,\Delta\theta_n) > q) \\ 
    & = 1 - P(\Delta\theta_i > q \hspace{.1cm} \forall i \in \{1,...,n\}) \\
    & = 1 - P(\Delta\theta_1>q,...,\Delta\theta_n>q) \\ 
    & \approx 1 - P(\Delta\theta_1>q)...P(\Delta\theta_n>q) \\ 
    & = 1 - P(\Delta\theta_1>q)^n,
\end{align*}
where the simplification in the second last line is an approximation due to unmodeled correlations in $\Delta\theta$s (though we expect it to be very good for large $n$). Thus the CDF of $Q_1$ is 
\[   
F_{Q_1}(q) = 
\left\{
\begin{array}{ll}
      0 & q < 0 \\
      1 - (\frac{\pi-q}{\pi})^n & q \in [0,\pi]  \\
      1 & q > \pi  \\
\end{array} 
\right., \]
which results in a PDF of 
\[   
f_{Q_1}(q) = \dfrac{d}{dy}{F_{Q_1}(q)} = 
\left\{
\begin{array}{ll}
      n\frac{(\pi-q)^{n-1}}{\pi^n} & ,\  q \in (0,\pi]  \\
      0 & ,\  \textrm{otherwise}  \\
\end{array} 
\right.. \]
Therefore the expectation of $Q_1$ is 
\begin{align*}
    E(Q_1) & = \int_{-\infty}^{\infty} q f_{Q_1}(q) dq  \\ 
           & = \int_{0}^{\pi} q n\frac{(\pi-q)^{n-1}}{\pi^n} dq \\ 
           & = \frac{\pi}{n+1}.
\end{align*}

Taking the $\Delta\theta$'s as independent, identically distributed random variables, when we look for $Q_2$, the $2^{nd}$ minimum, it is equivalent to taking the minimum of a set of $n-1$ $\Delta\theta$'s. By induction
\[E(Q_i) =  \frac{\pi}{n-i+2}.\]

Note that $n \neq N$ ($n$ is the number of pairs that can interact), and in the weak interaction case ($d \ll \sigma_r$), $n \sim N^2d/\sigma_r $. This is because, thinking of the random selection process as first choosing one particle and then randomly choosing a second one within the interaction range, we get $n \sim (N) (N d/\sigma_r)$.
Therefore 
\begin{equation}\label{eq:weak_interaction_theta}
    \mathbf{E}(\Delta\theta_\textrm{min}) \propto \mathbf{E}(Q_1) \propto \sigma_r/N^2d.
\end{equation}

In the strong interaction case $n \sim N^2$. This is because all particles have equal chance to collide with each other, so
\begin{equation*}
    n \sim \binom{N}{2} \sim N^2 .  
\end{equation*}
Therefore 
\begin{equation}\label{eq:strong_interaction_theta}
    \mathbf{E}(\Delta\theta_\textrm{min}) \propto 1/N^2.
\end{equation}

\subsection{Coupling strength scaling summary}
To summarize, in the weak interaction case, by Eqns.~\eqref{eq:weak_interaction_omega} and \eqref{eq:weak_interaction_theta}, we have
\begin{equation} \label{eq:weak_Kpw}
    \Kpw = \frac{\mathbf{E}(\Delta \omega)}{\mathbf{E}(\Delta \theta)} \sim \frac{d/\mu_r}{\frac{\sigma_r}{N^2 d}} = \frac{N^2 d^2}{\sigma_r \mu_r}.
\end{equation}

In the strong interaction case, by Eqns.~\eqref{eq:strong_interaction_omega} and \eqref{eq:strong_interaction_theta}, we have
\begin{equation} \label{eq:strong_Kpw}
    \Kpw = \frac{\mathbf{E}(\Delta \omega)}{\mathbf{E}(\Delta \theta)} \sim \frac{\sigma_r/\mu_r}{\frac{1}{N^2}} = \frac{N^2 \sigma_r}{ \mu_r}.
\end{equation}

\begin{figure}[h!]
    \centering
    \subfigure{\includegraphics[width=0.35\columnwidth]{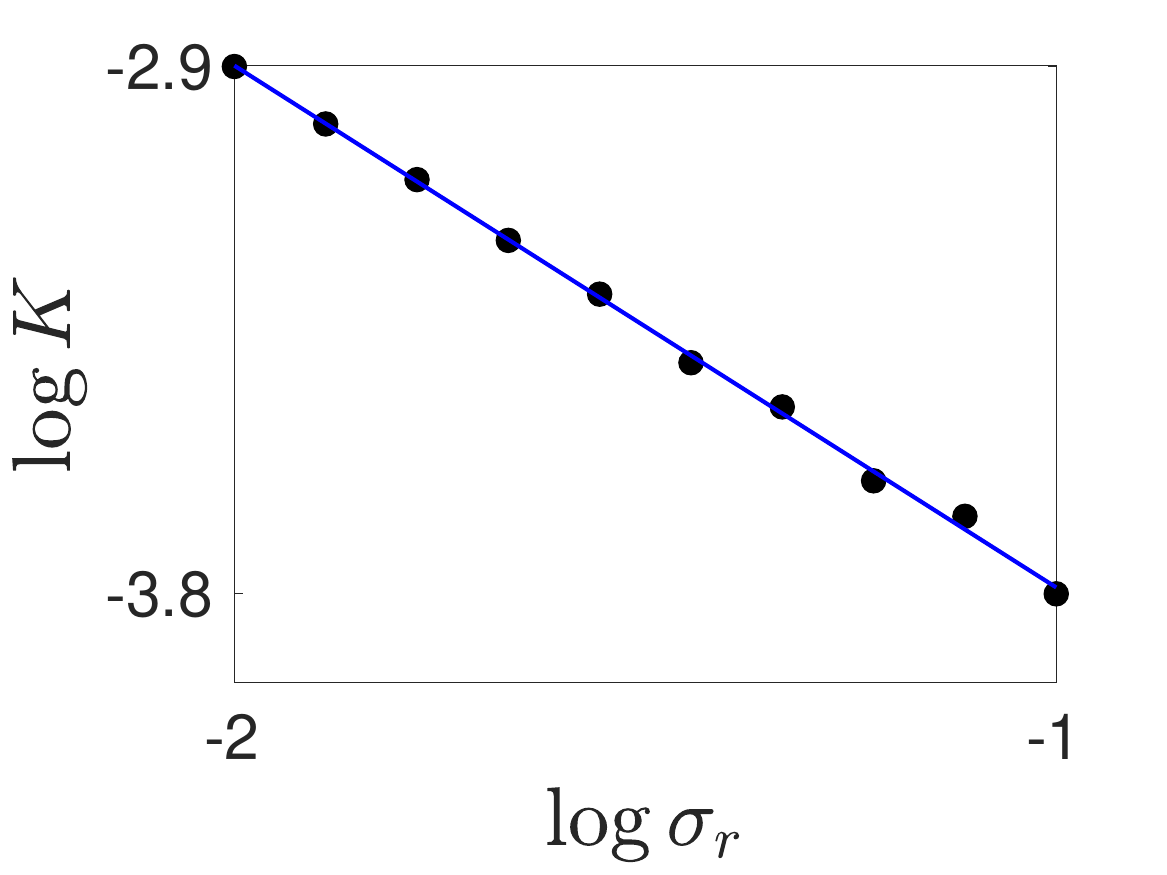}} 
    \subfigure{\includegraphics[width=0.35\columnwidth]{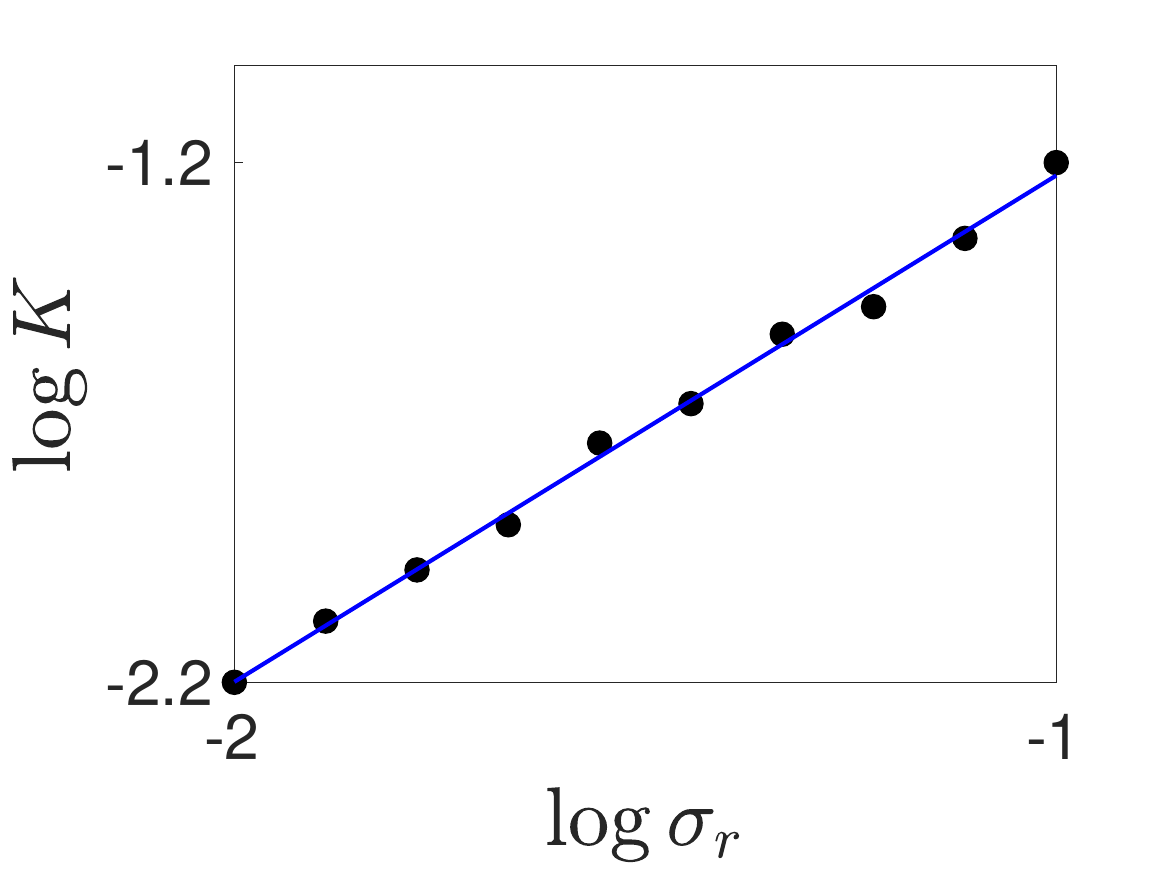}} 
    \caption{\textbf{Scaling of $\sigma_r$: weak interaction v.s. strong interaction.} The left panel shows the scaling of K versus $\sigma_r$ in the weak interaction case ($\sigma_r \gg d$, here we choose $\sigma_r \in (0.01, 0.1), d = 1$). The right panel shows the scaling of K versus $\sigma_r$ in the strong interaction case ($\sigma_r \sim d$, here we choose $\sigma_r \in (0.01, 0.1), d = 0.01$). The dots are from the same simulations as shown in Fig.~\ref{fig:couplingstrength} and the lines are linear fits. The slopes of the lines are consistent with the theoretical predictions for the scaling. 
    }
    \label{fig:scaling_strongvsweak}
\end{figure}

\begin{figure}[h!]
    \centering
    \subfigure{\includegraphics[width=0.35\columnwidth]{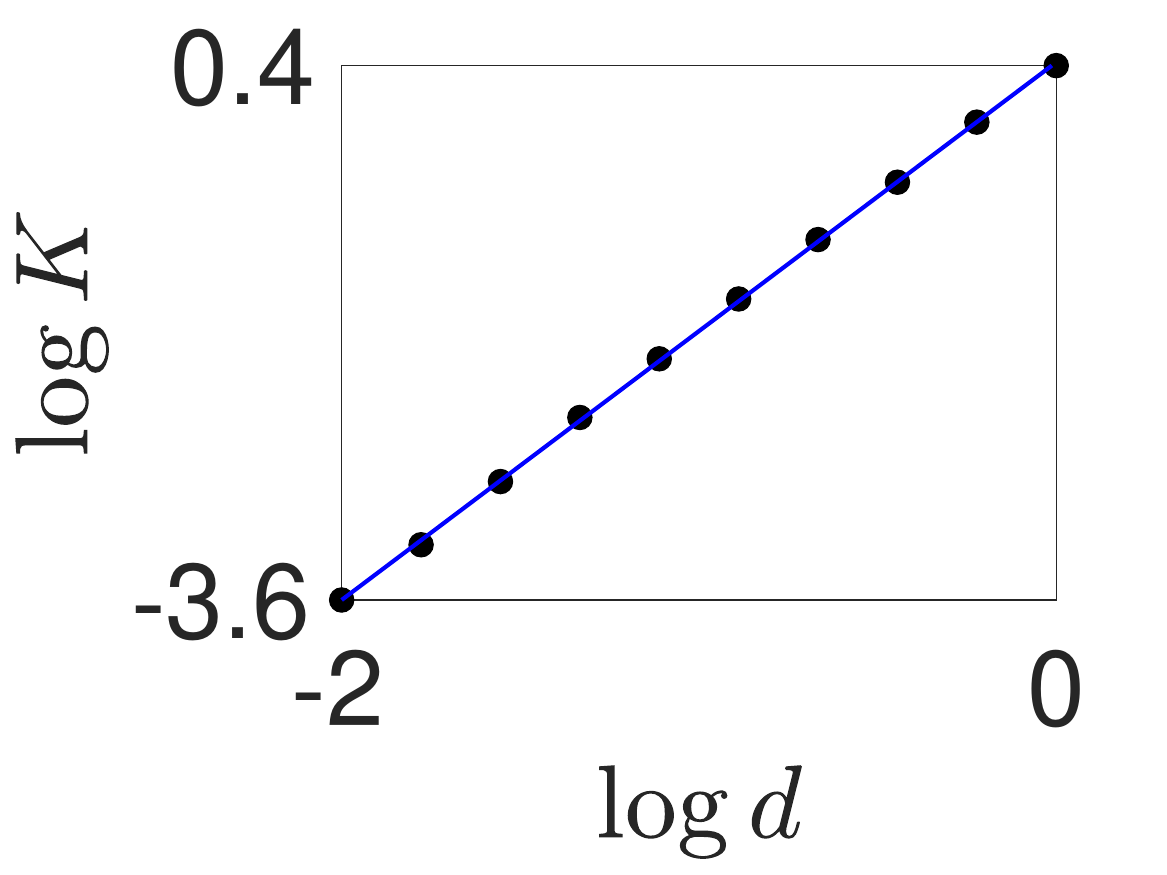}} 
    \subfigure{\includegraphics[width=0.35\columnwidth]{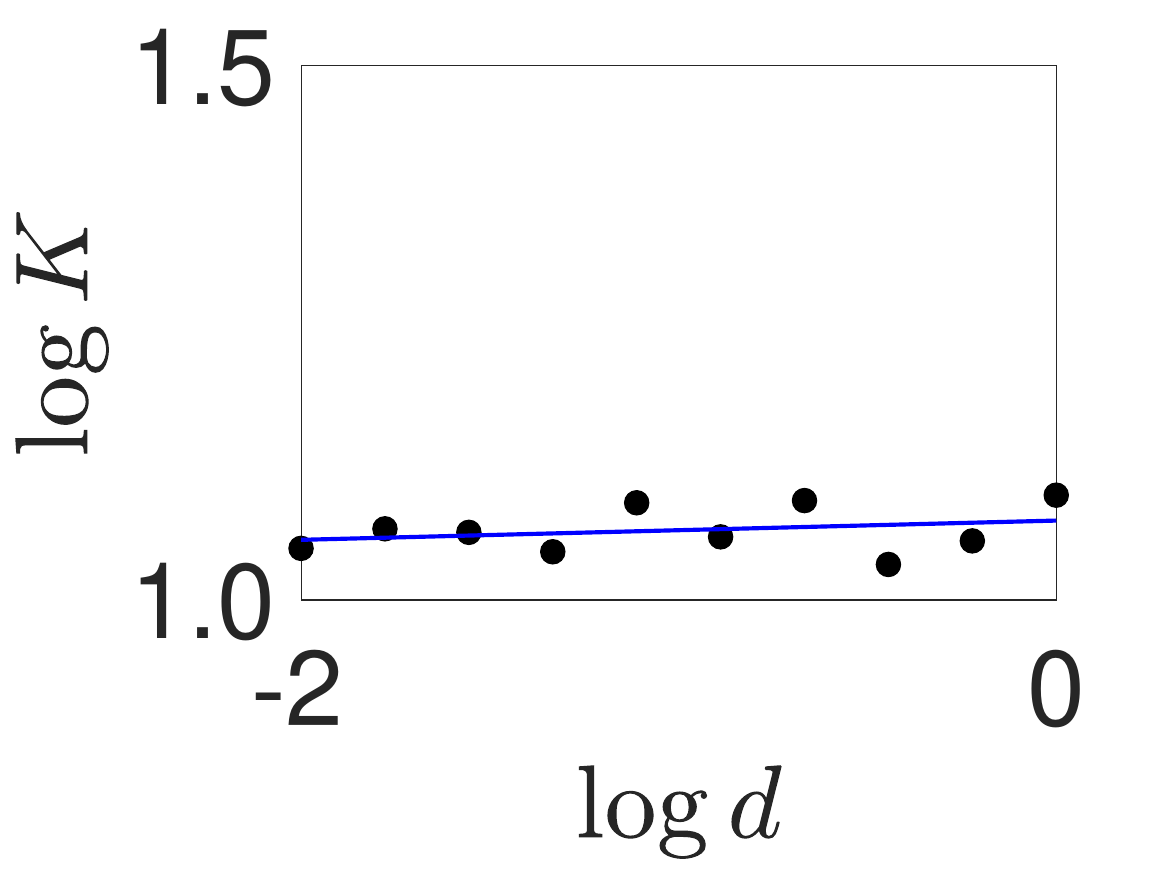}} 
    \caption{\textbf{Scaling of $d$: weak interaction v.s. strong interaction.} The left panel shows the scaling of K versus $d$ in the weak interaction case ($\sigma_r \gg d$). The right panel shows the scaling of K versus $d$ in the strong interaction case ($\sigma_r \sim d$. The dots are from simulations same as Fig.~\ref{fig:couplingstrength} and lines are the linear fits. The slope of the lines are consistent with the theoretical prediction of the scaling. 
    }
    \label{fig:scaling_strongvsweak_d}
\end{figure}

This scaling is consistent with the numerical results. The different scalings with respect to $\sigma_r$ and $d$ in the weak- and strong-interaction regimes are shown in Figs.~\ref{fig:scaling_strongvsweak} and \ref{fig:scaling_strongvsweak_d}, respectively.

\section{Optical depth-based argument for coupling strength scaling}

\subsection{Coupling strength scaling argument: 2D}\label{sec:2D_scalingargument}

In this section we relate the collision rate of orbiting particles to the optical depth of the system through a simple geometric scaling argument.  

Optical depth $\tau$ measures the total particle cross-sectional area projected onto the ring plane relative to the total area covered by the ring system. Interpreting $\tau$ as the expected number of particle cross sections encountered along one orbital circuit (divided by the orbital circuit area $\approx 2 \pi r_\textrm{orb} d$), a particle completes one orbit in time $\Torb$ and experiences, on average, $\tau$ collisions per orbit. For example, if the total (overlapping) cross-sectional area is twice the ring area ($\tau=2$), then the expected number of collisions in one orbital period is $2$, so the expected time between collisions is $\Torb/2$.

Accordingly, the expected time between collisions $\textrm{E}(\Tcoll)$ for a given particle is given by $\Torb / \tau$, where $\Torb = 2 \pi / \omega$ is the particle's orbital period and $\tau$ is the optical depth in the system 
\[ 
    \tau = \frac{(\textrm{number of particles})(\textrm{area of each particle})}{\textrm{area covered by ring system}}.
\]
So $\tau \sim N d^2 / (\sigma_r \mu_r)$. Note $\omega = (GM)^{1/2}\mu_r^{-3/2}$, and thus 
\[ 
    \textrm{E}(\Tcoll) \sim \frac{ (GM)^{-1/2}\mu_r^{5/2}\sigma_r}{N d^2}.
\]
The expected time until \textit{any} pair of particles collides is $\textrm{E}(\Tcoll) / N$ (shortened by a factor of $N$), which is therefore proportional to $\sigma_r / (N^2 d^2)$. Inter-collision time is the inverse of collision rate, which is linear with $\Kpw$, the pairwise coupling strength\footnote{This follows from viewing each collision as an ``interaction event'' that updates a particle's angular velocity in response to another particle. If the expected collision rate doubles, then the number of such interaction events per unit time also doubles, so the cumulative influence of neighbors on the phase dynamics over a fixed time window increases proportionally. Therefore, the coupling strength scales linearly with the collision rate.}. So we have 
\[
  \Kpw \propto \frac{N^2d^2}{\sigma_r(GM)^{-1/2}\mu_r^{5/2}}.
\]
This is consistent with results of numerical simulation. By rescaling the orbital period to $\Torb = 1$ (recall $\omega \propto \mu_r^{-3/2}$), we get
\begin{equation} \label{eq:Kpw_2D}
    \Kpw \propto \frac{N^2d^2}{\sigma_r\mu_r}.
\end{equation}
This quantity is dimension free, and is consistent with the weak interaction limit prediction above in Eq.~\eqref{eq:weak_Kpw}.

We note that, in this argument, we have assumed that each particle interaction contributes equally to the coupling---regardless of separation.  This is most consistent with the weak interaction limit considered in Section \ref{sec:Couplingstrengthscalingargument} above.  In the strong interaction limit, where even particles with quite disparate orbital radii can interaction, we expect that one would need to weight those interactions differently based on initial orbital radius disparity.

\section{With and without aggregation}\label{sec:withandwithout_aggregation}
\begin{figure}[t!]
    \centering
    \subfigure[]{\includegraphics[width=0.35\columnwidth]{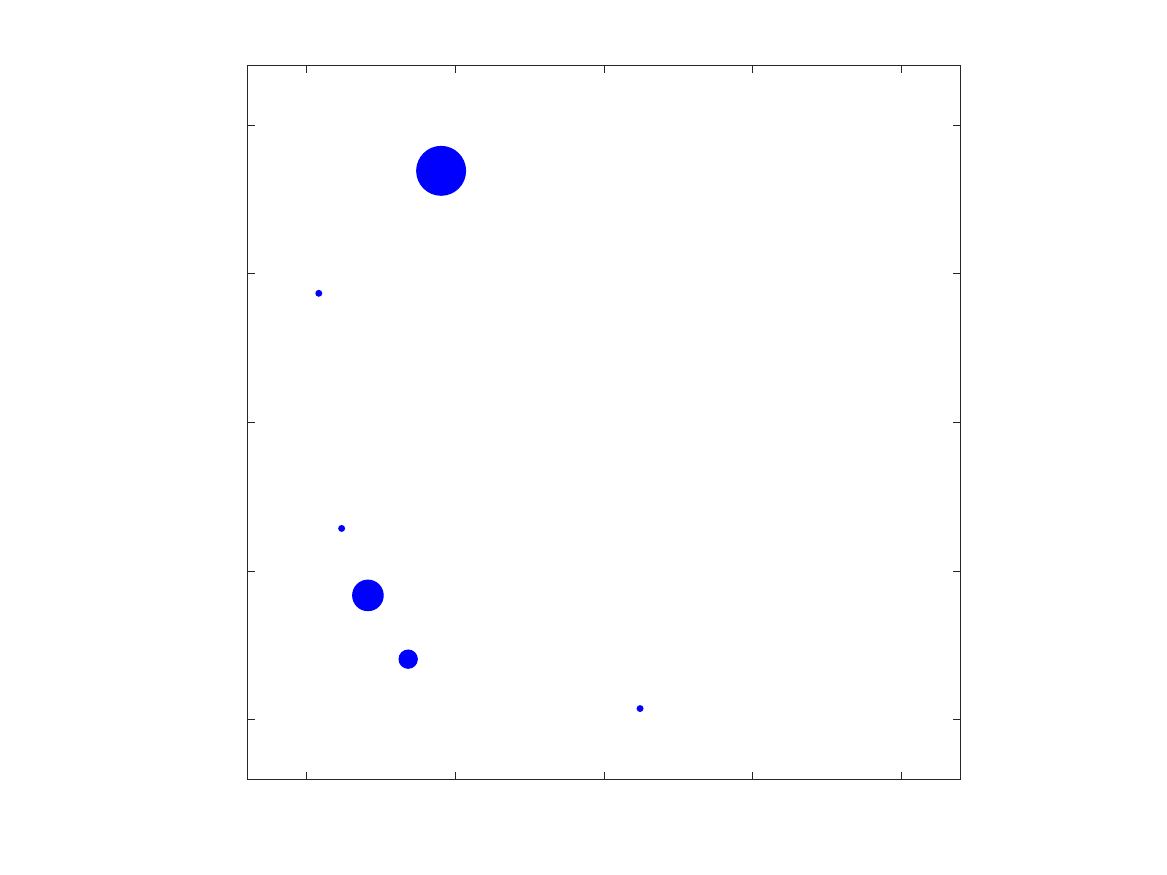}} 
    \subfigure[]{\includegraphics[width=0.35\columnwidth]{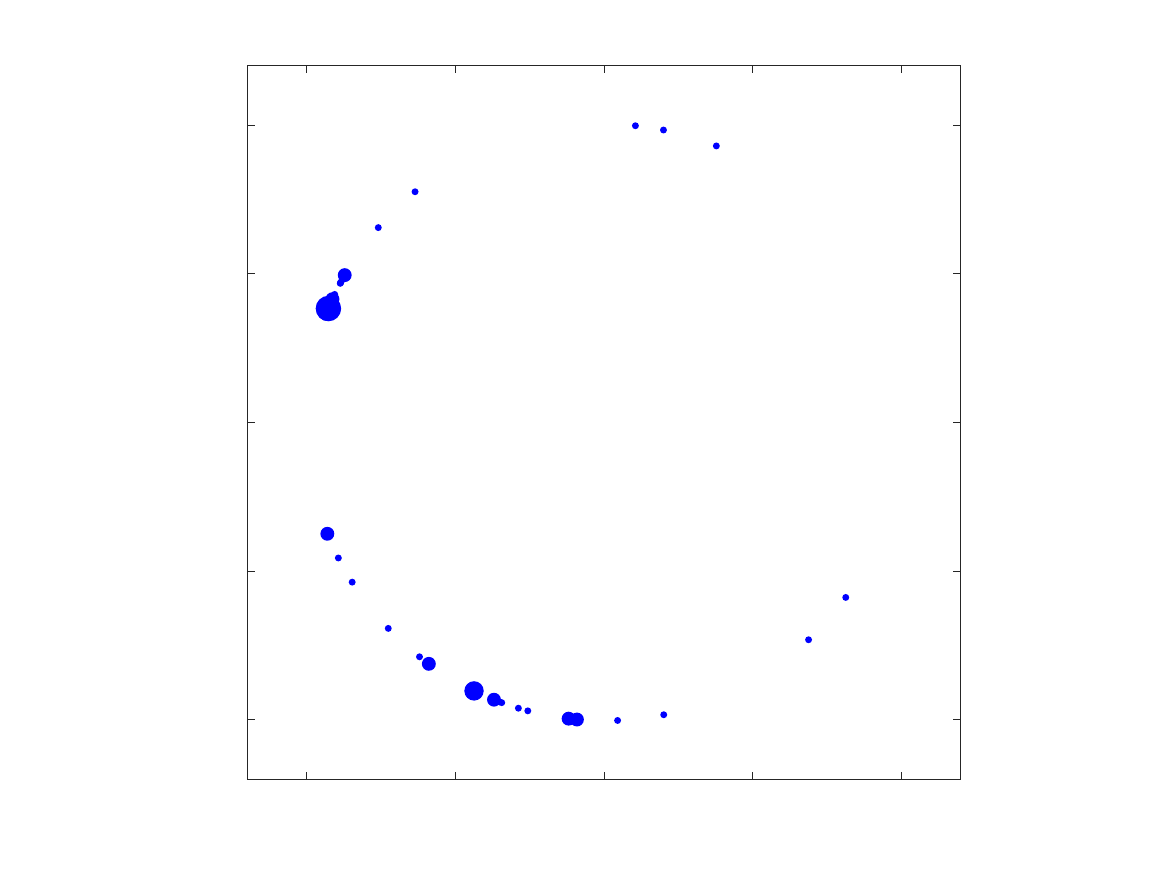}} 
    \caption{\textbf{Simulation results w/o aggregation.} (a) with aggregation (b) without aggregation. Parameters for simulation: $N=100, \mu_r = 1, \sigma_r = 0.02, d = 0.01, T = 10^5$. Aggregation distance $d_{a} = 10^{-6}$.}
    \label{fig:simulationaggregation}
\end{figure}

In Fig.~\ref{fig:simulationaggregation}, we run simulations with the same parameter set, comparing cases with and without aggregation. In the aggregation case, we allow particles to merge at each collision event if, after the collision, their orbital separation falls within the aggregation range. In the non-aggregation case, particles never merge during the dynamics. For visualization only, we group particles in the final plots if their orbits are closer than the aggregation range. The two sets of results are only qualitatively similar, since the outcomes are path dependent.

\section{Network structure:}
\begin{figure}
    \centering
    {\subfigure{\includegraphics[width=0.35\columnwidth, trim={30 30 0 30}, clip]{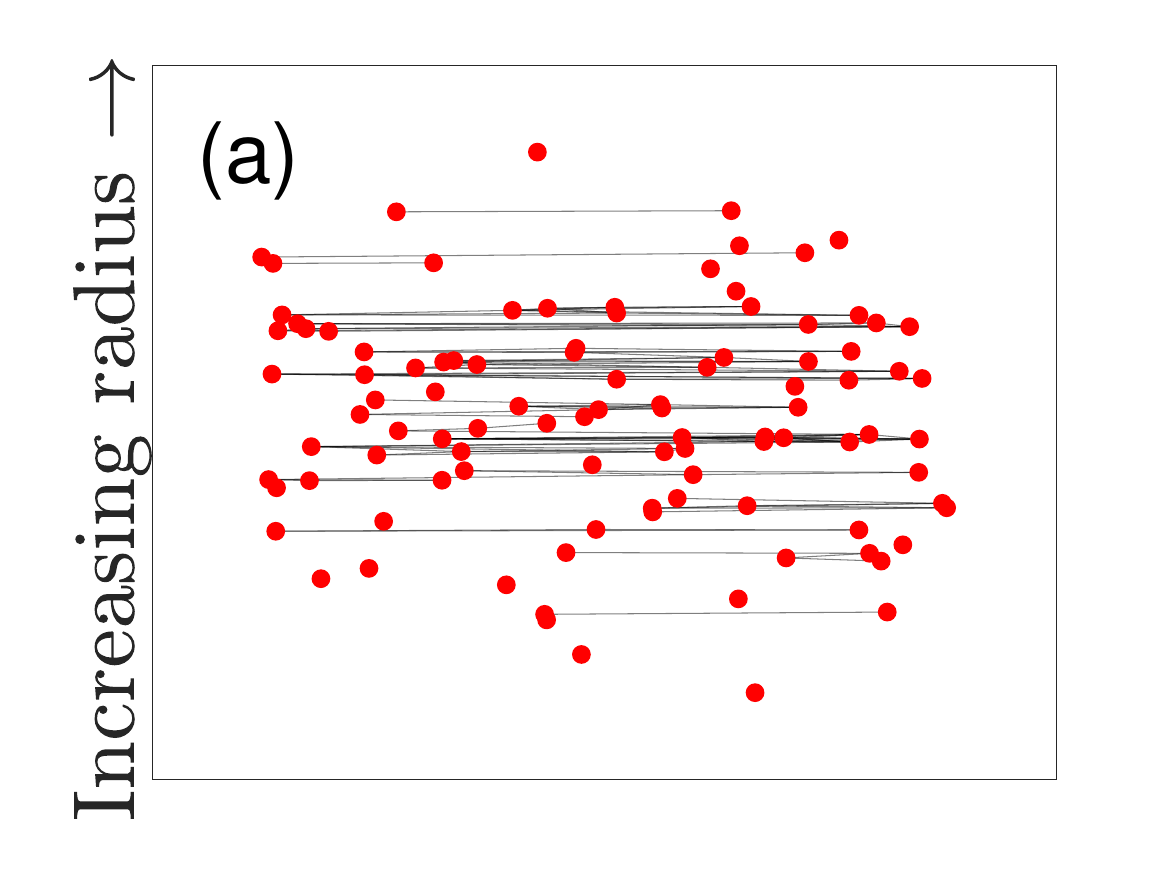}}}
    {\subfigure{\includegraphics[width=0.35\columnwidth, trim={30 30 0 30}, clip]{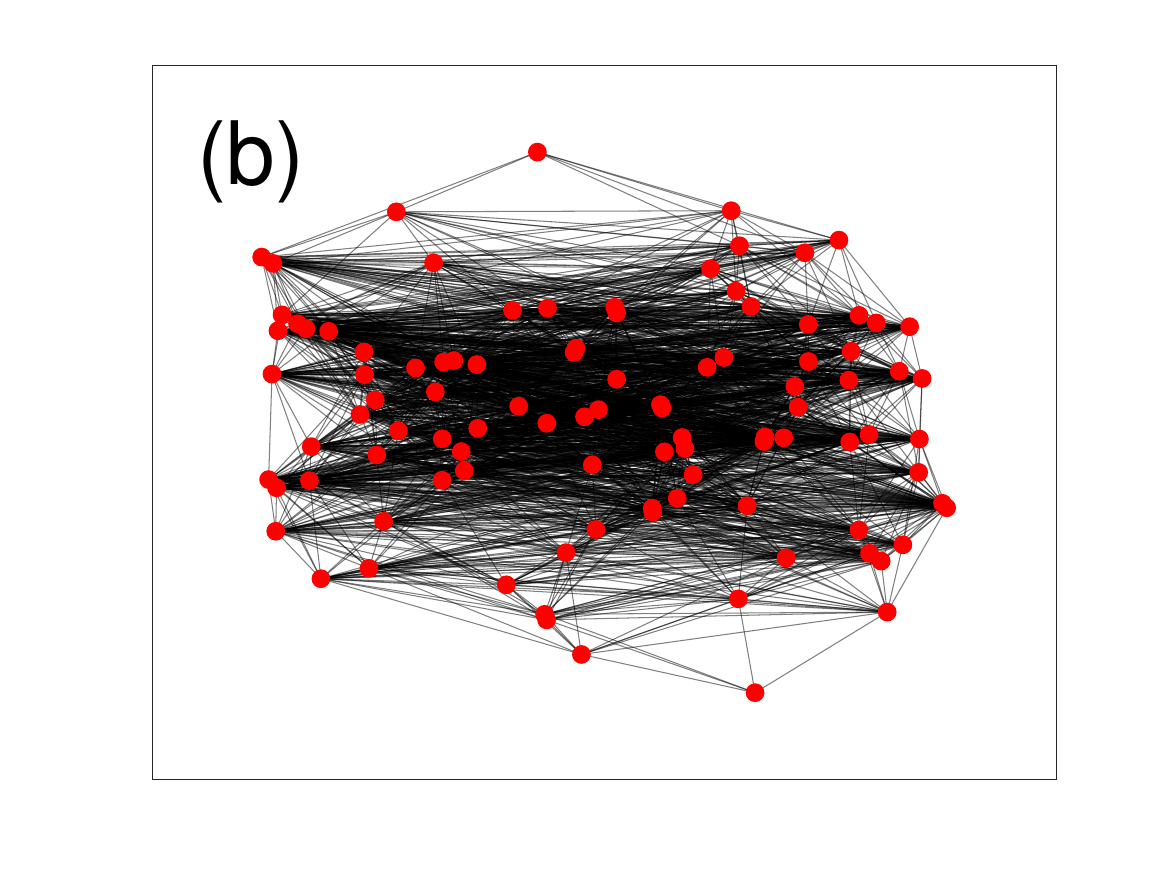}}}
    \caption{\textbf{Initial network structure for particles.} Here nodes represent particles and are connected if they are within the interaction range $d$.  In panel (a), $d / \sigma_r = 0.05$, so most nodes are disconnected.  In panel (b), $d / \sigma_r = 1$, so most nodes are connected.  Network layout is organized by particle orbital radius in the vertical direction and randomly in the horizontal direction.  Here $N=100$ and the radial distribution is Gaussian (see text). 
    }
    \label{fig:network}
\end{figure}

In general not every pair of particles can interact in our particle simulation, since for some pairs $|r_i - r_j| > d$).  Thus, in order to make more direct comparisons between the two models, we consider a variant of the KM with interactions restricted to a specified network:
\begin{equation} \label{eq:sawnetwork}
        \frac{d\theta_i}{dt} = \omega_i + \frac{K}{N}\sum_{j=1}^N D_{ij}\sin(\theta_j - \theta_i)\,,
\end{equation}
where
\begin{equation*}
D_{ij} = 
\left\{
\begin{array}{ll}
      1,&  |r_i-r_j| < d \\
      0,&  \textrm{otherwise}
\end{array} .
\right.
\end{equation*} 
The adjacency matrix elements $D_{ij}$ implicitly depends on the ratio $d / \sigma_r$ as $|r_i-r_j|$ depends on $\sigma_r$, the ratio of interaction range to standard deviation of particle orbital radius. In Fig.~\ref{fig:network}, we visualize the network structure for two different values of $d / \sigma_r$: when $d / \sigma_r$ is small, we observe chain-like sparse networks, but when $d / \sigma_r$ is large, we get dense networks. Note that we used a Gaussian distribution for initial particle radii in the figure; the network structure can be very different with other distributions. 

\section{Exact post-collision radii calculation} \label{SM:exact_polynomial}
Solving Eqns.~\eqref{eq:momentum} and \eqref{eq:energy} with $\epsilon'=0$ while restricting to $0 \leq \epsilon < 1$ and $m_1, m_2,\rione, \ritwo, \rfone, \rftwo > 0$ yields a quartic polynomial to be solved for the post-collision radii $z$:  
\begin{equation} \label{eq:collision}
    C_4z^4 + C_3z^3 + C_2z^2 + C_1z + C_0 ,
\end{equation}	
where the coefficients are:
\scriptsize
\begin{align*}
    C_4 = &\epsilon^{3} m_1^{3} r_2^{3} r_3^{2}
    +3 \epsilon^{3} m_1^{2} m_2 r_1 \, r_2^{2} r_3^{2}
    +3 \epsilon^{3} m_1 \, m_2^{2} r_1^{2} r_2 \, r_3^{2}
    +\epsilon^{3} m_2^{3} r_1^{3} r_3^{2} \\
    & -4 \epsilon^{2} m_1^{3} r_1 \, r_2^{3} r_3
    -\epsilon^{2} m_1^{3} r_2^{3} r_3^{2}
    -8 \epsilon^{2} m_1^{2} m_2 \, r_1^{2} r_2^{2} r_3
    -3 \epsilon^{2} m_1^{2} m_2 r_1 \, r_2^{2} r_3^{2} \\
    & -4 \epsilon^{2} m_1 \, m_2^{2} r_1^{3} r_2 r_3
    -3 \epsilon^{2} m_1 \, m_2^{2} r_1^{2} r_2 \, r_3^{2}
    -\epsilon^{2} m_2^{3} r_1^{3} r_3^{2}
    +4 \epsilon \, m_1^{3} r_1^{2} r_2^{3} \\
    & +4 \epsilon \, m_1^{3} r_1 \, r_2^{3} r_3
    +4 \epsilon \, m_1^{2} m_2 \, r_1^{3} r_2^{2}
    +8 \epsilon \, m_1^{2} m_2 \, r_1^{2} r_2^{2} r_3
    +4 \epsilon \, m_1 \, m_2^{2} r_1^{3} r_2 r_3 \\
    & -4 m_1^{3} r_1^{2} r_2^{3}
    -4 m_1^{2} m_2 \, r_1^{3} r_2^{2}
    \end{align*}

\begin{align*}
    C_3 = &-8 r_1^{\frac{5}{2}} \epsilon \, m_1^{3} r_2^{3}
    +4 r_3^{\frac{5}{2}} \epsilon^{2} m_1^{3} r_2^{3}
    +4 r_3^{\frac{5}{2}} \epsilon^{2} m_2^{3} r_1^{3}
    -4 r_3^{\frac{5}{2}} \epsilon^{3} m_1^{3} r_2^{3} \\
    & -4 r_3^{\frac{5}{2}} \epsilon^{3} m_2^{3} r_1^{3}
    -12 r_3^{\frac{5}{2}} \epsilon^{3} m_1^{2} m_2 r_1 \, r_2^{2}
    -12 r_3^{\frac{5}{2}} \epsilon^{3} m_1 \, m_2^{2} r_1^{2} r_2 \\
    & +4 r_1^{\frac{7}{2}} \epsilon^{2} m_1 \, m_2^{2} r_2 r_3
    +4 r_2^{\frac{7}{2}} \epsilon^{2} m_1^{2} m_2 r_1 r_3
    +8 r_1^{\frac{5}{2}} \epsilon^{2} m_1^{2} m_2 \, r_2^{2} r_3 \\
    & +8 r_2^{\frac{5}{2}} \epsilon^{2} m_1 \, m_2^{2} r_1^{2} r_3
    +16 r_3^{\frac{3}{2}} \epsilon^{2} m_1^{2} m_2 \, r_1^{2} r_2^{2}
    +8 r_3^{\frac{3}{2}} \epsilon^{2} m_1 \, m_2^{2} r_1^{3} r_2 \\
    & -8 r_1^{\frac{7}{2}} \epsilon \, m_1^{2} m_2 \, r_2^{2}
    -8 r_2^{\frac{7}{2}} \epsilon \, m_1^{2} m_2 \, r_1^{2}
    +8 r_3^{\frac{3}{2}} \epsilon^{2} m_1^{3} r_1 \, r_2^{3}
    +4 r_1^{\frac{3}{2}} \epsilon^{2} m_1^{3} r_2^{3} r_3 \\
    & -8 r_2^{\frac{5}{2}} \epsilon \, m_1 \, m_2^{2} r_1^{3}
    +4 r_2^{\frac{3}{2}} \epsilon^{2} m_2^{3} r_1^{3} r_3
    +8 m_1 \, m_2^{2} r_1^{3} r_2^{\frac{5}{2}}
    +8 r_1^{\frac{7}{2}} m_1^{2} m_2 \, r_2^{2} \\
    & +8 m_1^{2} m_2 \, r_1^{2} r_2^{\frac{7}{2}}
    +8 r_1^{\frac{5}{2}} m_1^{3} r_2^{3}
    -8 r_3^{\frac{3}{2}} \epsilon \, m_1^{3} r_1 \, r_2^{3}
    -4 r_1^{\frac{3}{2}} \epsilon \, m_1^{3} r_2^{3} r_3 \\
    & -4 r_2^{\frac{3}{2}} \epsilon \, m_2^{3} r_1^{3} r_3
    +12 r_3^{\frac{5}{2}} \epsilon^{2} m_1^{2} m_2 r_1 \, r_2^{2}
    +12 r_3^{\frac{5}{2}} \epsilon^{2} m_1 \, m_2^{2} r_1^{2} r_2 \\
    & -4 r_1^{\frac{7}{2}} \epsilon \, m_1 \, m_2^{2} r_2 r_3
    -4 m_1^{2} m_2 \, r_1 \, r_2^{\frac{7}{2}} r_3 \epsilon
    -8 r_1^{\frac{5}{2}} \epsilon \, m_1^{2} m_2 \, r_2^{2} r_3 \\
    & -8 r_2^{\frac{5}{2}} \epsilon \, m_1 \, m_2^{2} r_1^{2} r_3
    -16 r_3^{\frac{3}{2}} \epsilon \, m_1^{2} m_2 \, r_1^{2} r_2^{2}
    -8 r_3^{\frac{3}{2}} \epsilon \, m_1 \, m_2^{2} r_1^{3} r_2
\end{align*}

\begin{align*}
    C_2 = &-12 \epsilon^{2} m_1^{2} m_2 r_1 \, r_2^{2} r_3^{3}
    +3 \epsilon^{2} m_1 \, m_2^{2} r_1^{3} r_2 \, r_3^{2}
    +6 \epsilon^{2} m_1 \, m_2^{2} r_1^{2} r_2^{2} r_3^{2} \\
    & -12 \epsilon^{2} m_1 \, m_2^{2} r_1^{2} r_2 \, r_3^{3}
    -6 \epsilon \, m_1^{2} m_2 \, r_1^{3} r_2^{2} r_3
    -6 \epsilon \, m_1 \, m_2^{2} r_1^{2} r_2^{3} r_3 \\
    & -8 r_3^{\frac{3}{2}} r_1^{\frac{7}{2}} \epsilon^{2} m_1 \, m_2^{2} r_2
    -8 r_3^{\frac{3}{2}} r_2^{\frac{7}{2}} \epsilon^{2} m_1^{2} m_2 r_1
    +8 r_3^{\frac{3}{2}} r_1^{\frac{7}{2}} \epsilon \, m_1 \, m_2^{2} r_2 \\
    & -16 r_3^{\frac{3}{2}} r_1^{\frac{5}{2}} \epsilon^{2} m_1^{2} m_2 \, r_2^{2}
    +8 r_3^{\frac{3}{2}} r_2^{\frac{7}{2}} \epsilon \, m_1^{2} m_2 r_1
    -16 r_3^{\frac{3}{2}} r_2^{\frac{5}{2}} \epsilon^{2} m_1 \, m_2^{2} r_1^{2} \\
    & +16 r_3^{\frac{3}{2}} r_1^{\frac{5}{2}} \epsilon \, m_1^{2} m_2 \, r_2^{2}
    +16 r_3^{\frac{3}{2}} r_2^{\frac{5}{2}} \epsilon \, m_1 \, m_2^{2} r_1^{2}
    +3 \epsilon^{3} m_1^{3} r_2^{3} r_3^{3} \\
    & +3 \epsilon^{3} m_2^{3} r_1^{3} r_3^{3}
    -4 \epsilon^{2} m_1^{3} r_2^{3} r_3^{3}
    -4 \epsilon^{2} m_2^{3} r_1^{3} r_3^{3}
    +4 \epsilon \, m_1^{3} r_1^{3} r_2^{3} \\
    & +4 \epsilon \, m_2^{3} r_1^{3} r_2^{3}
    -4 m_1^{2} m_2 \, r_1^{4} r_2^{2}
    -4 m_1 \, m_2^{2} r_1^{2} r_2^{4}
    -8 r_1^{\frac{7}{2}} r_2^{\frac{5}{2}} m_1 \, m_2^{2} \\
    & -8 r_1^{\frac{5}{2}} r_2^{\frac{7}{2}} m_1^{2} m_2
    +3 \epsilon^{2} m_1^{3} r_1 \, r_2^{3} r_3^{2}
    +3 \epsilon^{2} m_2^{3} r_1^{3} r_2 \, r_3^{2}
    -6 \epsilon \, m_1^{3} r_1^{2} r_2^{3} r_3 \\
    & +4 \epsilon \, m_1^{2} m_2 \, r_1^{4} r_2^{2}
    +4 \epsilon \, m_1 \, m_2^{2} r_1^{2} r_2^{4}
    -6 \epsilon \, m_2^{3} r_1^{3} r_2^{2} r_3
    +8 r_1^{\frac{7}{2}} r_2^{\frac{5}{2}} \epsilon \, m_1 \, m_2^{2} \\
    & +8 r_1^{\frac{5}{2}} r_2^{\frac{7}{2}} \epsilon \, m_1^{2} m_2
    -8 r_3^{\frac{3}{2}} r_1^{\frac{3}{2}} \epsilon^{2} m_1^{3} r_2^{3}
    -8 r_3^{\frac{3}{2}} r_2^{\frac{3}{2}} \epsilon^{2} m_2^{3} r_1^{3} \\
    & +8 r_3^{\frac{3}{2}} r_1^{\frac{3}{2}} \epsilon \, m_1^{3} r_2^{3}
    +8 r_3^{\frac{3}{2}} r_2^{\frac{3}{2}} \epsilon \, m_2^{3} r_1^{3}
    +9 \epsilon^{3} m_1^{2} m_2 r_1 \, r_2^{2} r_3^{3} \\
    & +9 \epsilon^{3} m_1 \, m_2^{2} r_1^{2} r_2 \, r_3^{3}
    +6 \epsilon^{2} m_1^{2} m_2 \, r_1^{2} r_2^{2} r_3^{2}
    +3 \epsilon^{2} m_1^{2} m_2 r_1 \, r_2^{3} r_3^{2}
\end{align*}

\begin{align*}
    C_1 = &4 m_1^{3} r_1^{4} r_2^{3}
    -2 \epsilon \, m_1^{2} m_2 \, r_1^{4} r_2^{2} r_3
    -2 \epsilon \, m_1 m_2^{2} r_1^{2} r_2^{4} r_3 \\
    & +4 r_3^{\frac{5}{2}} r_1^{\frac{3}{2}} \epsilon^{2} m_1^{3} r_2^{3}
    +4 r_3^{\frac{5}{2}} r_2^{\frac{3}{2}} \epsilon^{2} m_2^{3} r_1^{3}
    +8 r_1^{\frac{7}{2}} r_2^{\frac{7}{2}} m_1^{2} m_2 \\
    & -8 r_3^{\frac{3}{2}} r_1^{\frac{5}{2}} \epsilon \, m_1^{3} r_2^{3}
    -6 \epsilon^{3} m_1^{2} m_2 r_1 \, r_2^{2} r_3^{4}
    -6 \epsilon^{3} m_1 \, m_2^{2} r_1^{2} r_2 \, r_3^{4} \\
    & +8 \epsilon^{2} m_1^{2} m_2 \, r_1^{2} r_2^{2} r_3^{3}
    +4 \epsilon^{2} m_1 \, m_2^{2} r_1^{3} r_2 \, r_3^{3}
    -8 r_3^{\frac{3}{2}} r_2^{\frac{5}{2}} \epsilon \, m_1 \, m_2^{2} r_1^{3} \\
    & +8 r_3^{\frac{5}{2}} r_1^{\frac{5}{2}} \epsilon^{2} m_1^{2} m_2 \, r_2^{2}
    +8 r_3^{\frac{5}{2}} r_2^{\frac{5}{2}} \epsilon^{2} m_1 \, m_2^{2} r_1^{2}
    -4 r_1^{\frac{5}{2}} r_2^{\frac{7}{2}} \epsilon \, m_1^{2} m_2 r_3 \\
    & -4 r_1^{\frac{7}{2}} r_2^{\frac{5}{2}} \epsilon \, m_1 \, m_2^{2} r_3
    -8 r_3^{\frac{3}{2}} r_1^{\frac{7}{2}} \epsilon \, m_1^{2} m_2 \, r_2^{2}
    -8 r_3^{\frac{3}{2}} r_2^{\frac{7}{2}} \epsilon \, m_1^{2} m_2 \, r_1^{2} \\
    & +4 m_1 \, m_2^{2} r_1^{3} r_2^{4}
    -2 \epsilon^{3} m_1^{3} r_2^{3} r_3^{4}
    -2 \epsilon^{3} m_2^{3} r_1^{3} r_3^{4} \\
    & +4 r_3^{\frac{5}{2}} r_1^{\frac{7}{2}} \epsilon^{2} m_1 \, m_2^{2} r_2
    +4 r_3^{\frac{5}{2}} r_2^{\frac{7}{2}} \epsilon^{2} m_1^{2} m_2 r_1
    +4 \epsilon^{2} m_1^{3} r_1 \, r_2^{3} r_3^{3} \\
    & -2 \epsilon \, m_1^{3} r_1^{3} r_2^{3} r_3
    -2 \epsilon \, m_2^{3} r_1^{3} r_2^{3} r_3
    \end{align*}
    \begin{align*}
    C_0 = &4 m_1^{3} r_1^{4} r_2^{3}
    -2 \epsilon \, m_1^{2} m_2 \, r_1^{4} r_2^{2} r_3
    -2 \epsilon \, m_1 m_2^{2} r_1^{2} r_2^{4} r_3 \\
    & +4 r_3^{\frac{5}{2}} r_1^{\frac{3}{2}} \epsilon^{2} m_1^{3} r_2^{3}
    +4 r_3^{\frac{5}{2}} r_2^{\frac{3}{2}} \epsilon^{2} m_2^{3} r_1^{3}
    +8 r_1^{\frac{7}{2}} r_2^{\frac{7}{2}} m_1^{2} m_2 \\
    & -8 r_3^{\frac{3}{2}} r_1^{\frac{5}{2}} \epsilon \, m_1^{3} r_2^{3}
    -6 \epsilon^{3} m_1^{2} m_2 r_1 \, r_2^{2} r_3^{4}
    -6 \epsilon^{3} m_1 \, m_2^{2} r_1^{2} r_2 \, r_3^{4} \\
    & +8 \epsilon^{2} m_1^{2} m_2 \, r_1^{2} r_2^{2} r_3^{3}
    +4 \epsilon^{2} m_1 \, m_2^{2} r_1^{3} r_2 \, r_3^{3}
    -8 r_3^{\frac{3}{2}} r_2^{\frac{5}{2}} \epsilon \, m_1 \, m_2^{2} r_1^{3} \\
    & +8 r_3^{\frac{5}{2}} r_1^{\frac{5}{2}} \epsilon^{2} m_1^{2} m_2 \, r_2^{2}
    +8 r_3^{\frac{5}{2}} r_2^{\frac{5}{2}} \epsilon^{2} m_1 \, m_2^{2} r_1^{2}
    -4 r_1^{\frac{5}{2}} r_2^{\frac{7}{2}} \epsilon \, m_1^{2} m_2 r_3 \\
    & -4 r_1^{\frac{7}{2}} r_2^{\frac{5}{2}} \epsilon \, m_1 \, m_2^{2} r_3
    -8 r_3^{\frac{3}{2}} r_1^{\frac{7}{2}} \epsilon \, m_1^{2} m_2 \, r_2^{2}
    -8 r_3^{\frac{3}{2}} r_2^{\frac{7}{2}} \epsilon \, m_1^{2} m_2 \, r_1^{2} \\
    & +4 m_1 \, m_2^{2} r_1^{3} r_2^{4}
    -2 \epsilon^{3} m_1^{3} r_2^{3} r_3^{4}
    -2 \epsilon^{3} m_2^{3} r_1^{3} r_3^{4} \\
    & +4 r_3^{\frac{5}{2}} r_1^{\frac{7}{2}} \epsilon^{2} m_1 \, m_2^{2} r_2
    +4 r_3^{\frac{5}{2}} r_2^{\frac{7}{2}} \epsilon^{2} m_1^{2} m_2 r_1
    +4 \epsilon^{2} m_1^{3} r_1 \, r_2^{3} r_3^{3} \\
    & -2 \epsilon \, m_1^{3} r_1^{3} r_2^{3} r_3
    -2 \epsilon \, m_2^{3} r_1^{3} r_2^{3} r_3 .
\end{align*}

\normalsize
In order to obtain physical solutions for the post-collision radii, we require at least two roots of Eq.~\eqref{eq:collision} to be real and positive (i.e., impose the critical condition that Eq.~\eqref{eq:collision} take the form $(z-z_0)^2(z-z_1)(z-z_2)$), leading to the inequality
\begin{equation} \label{eq:epsmax}
	\epsilon \leq \epsmax \approx \frac{f(r_1,r_2)}{r_3^{3/2}} + \mathcal{O}(r_3^{-2}),
\end{equation}
where $f(r_1,r_2) = \frac{r_1 r_2}{r_1 + r_2} \left(\sqrt{r_1}+\sqrt{r_2}-2\sqrt{\frac{2r_1 r_2}{r_1 + r_2}}\right)$. The explicit exact formula for $\epsmax$ is omitted here; approximation is obtained when $r_3 \gg r_i, i = 1,2$ and $m_1 = m_2 = m$.

Thus, conservation of energy (i.e., $\epsilon'=0$) and angular momentum impose an upper bound on the fraction of mass that can be lost per collision, which goes to zero as $r_1 \to r_2$. When there is mass loss in this scenario, the radial distance between two colliding bodies gets smaller.

\section{Implications for the fine structure of Saturn's rings}

\begin{figure}
    \centering
    {\subfigure{\includegraphics[width=0.45\columnwidth]{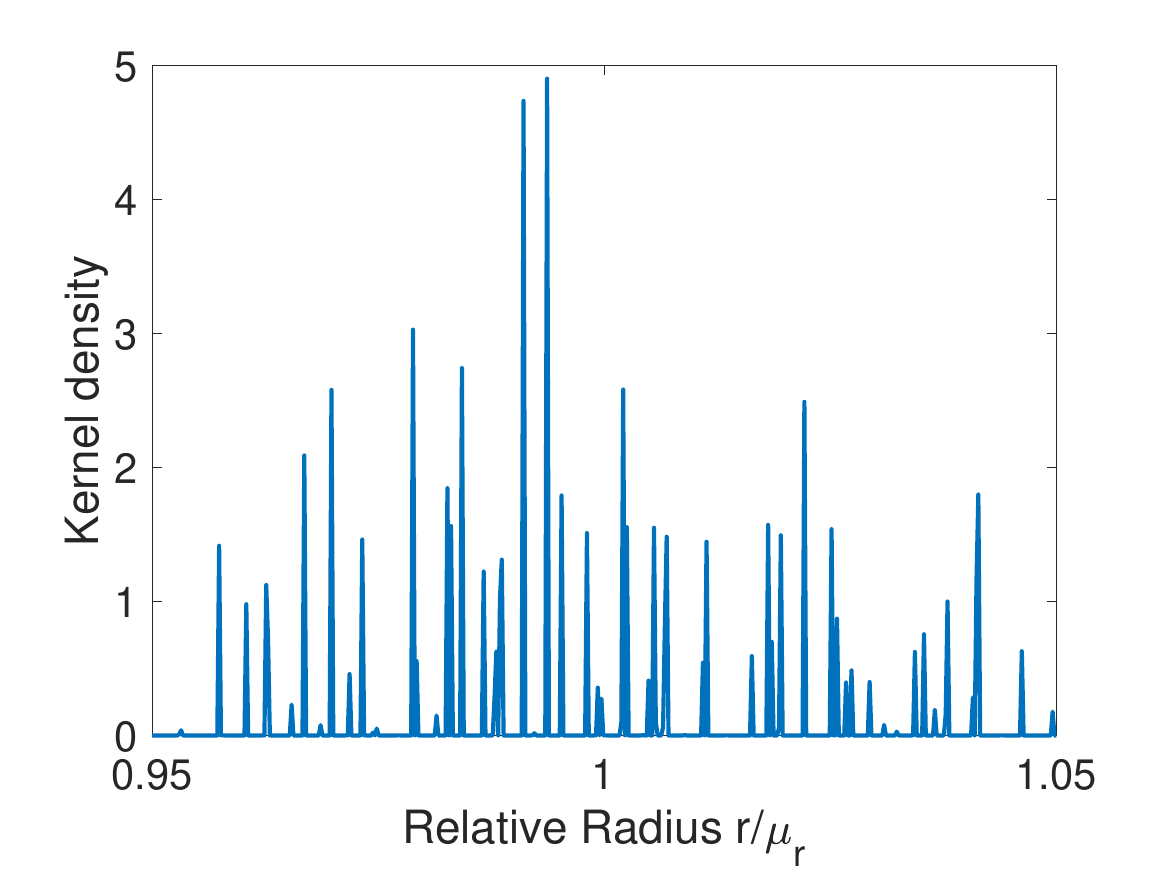}}}
    {\subfigure{\includegraphics[width=0.45\columnwidth]{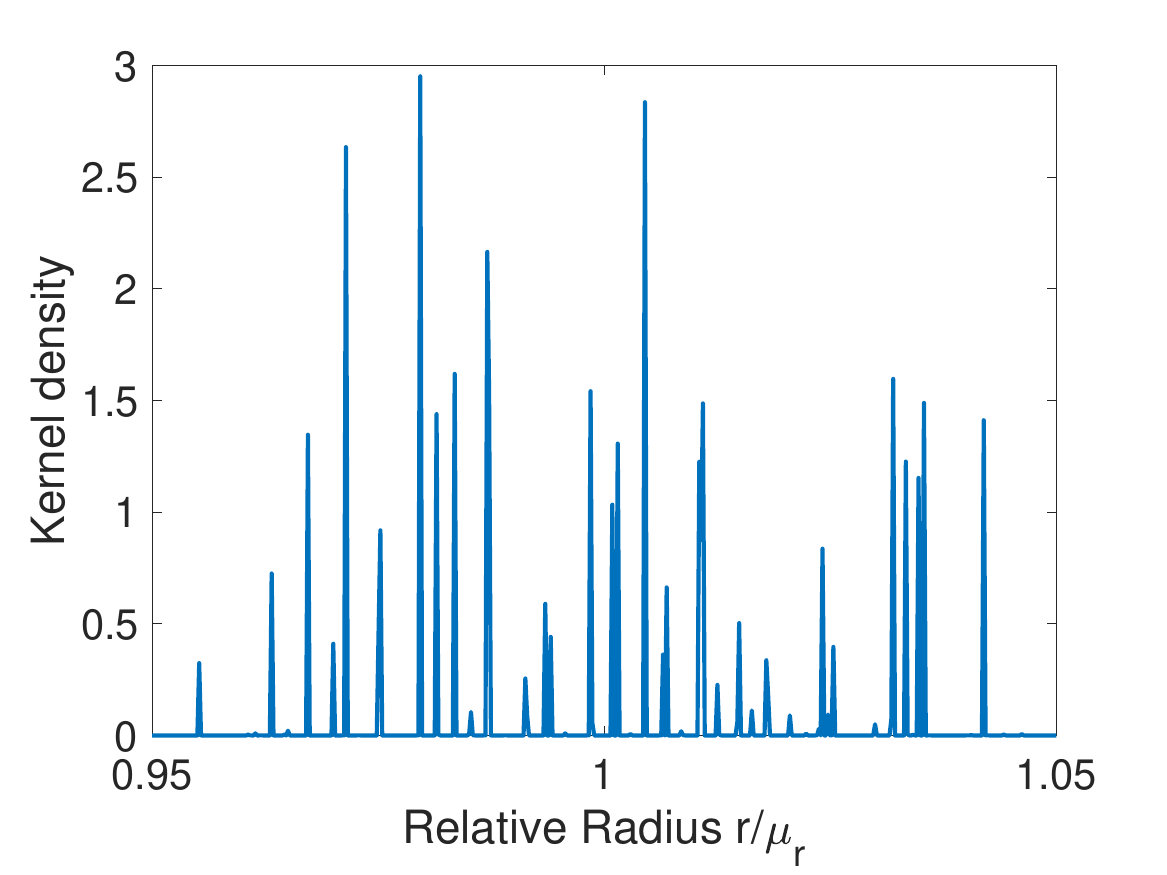}}}
    \caption{\textbf{Fine structure of rings. 
    }
    Kernel density estimates (KDEs) of the particle orbital-radius distribution. The left and right panels correspond to simulations with identical parameters but different random seeds, illustrating the sensitivity of the final ring structure to stochastic fluctuations. Simulations were performed with $N=400$, $d=0.1$, $\sigma_r=0.02$, $\mu_r=100$, and $T=5\times10^4$. The KDE bandwidth was set to $h=0.0005$.
    }
    \label{fig:finestructure}
\end{figure}

The collision model may also provide a new framework for understanding the fine radial structure observed in Saturn’s rings. Rather than forming smooth, continuous disks, Saturn’s rings exhibit a rich hierarchy of radial and azimuthal features, including narrow ringlets, gaps, and density variations \cite{Lehmann2017}. Figure~\ref{fig:finestructure} presents examples of final ring structures generated by our simulations. These examples demonstrate that an initially smooth particle distribution can evolve into localized radial structures through the collision dynamics alone. While these patterns should not be interpreted as direct reproductions of specific features observed in Saturn’s rings, they illustrate that the collision mechanism proposed here is capable of generating fine-scale spatial organization and may provide a plausible route for the emergence of such structures.

\end{document}